\documentclass[fleqn,usenatbib]{mnras}

\usepackage[T1]{fontenc}

\DeclareRobustCommand{\VAN}[3]{#2}
\let\VANthebibliography\thebibliography
\def\thebibliography{\DeclareRobustCommand{\VAN}[3]{##3}\VANthebibliography}

\hypersetup{linkcolor=blue,citecolor=blue,filecolor=cyan,urlcolor=magenta}
\usepackage{graphicx}   
\usepackage{amsmath}    
\usepackage{amssymb}    
\usepackage{newtxtext,newtxmath}
\usepackage{color}
\usepackage{threeparttable}
\usepackage{hyperref}

\newcommand{\lta}{\lower 2pt \hbox{$\, \buildrel {\scriptstyle <}\over {\scriptstyle \sim}\,$}}
\newcommand{\gta}{\lower 2pt \hbox{$\, \buildrel {\scriptstyle >}\over {\scriptstyle \sim}\,$}}
\definecolor{blazeorange}{rgb}{1.0, 0.4, 0.0}
\definecolor{seagreen}{rgb}{0.18, 0.55, 0.34}
\definecolor{rufous}{rgb}{0.66, 0.11, 0.03}
\definecolor{royalfuchsia}{rgb}{0.79, 0.17, 0.57}
\definecolor{scarlet}{rgb}{1.0, 0.13, 0.0}
\definecolor{royalpurple}{rgb}{0.47, 0.32, 0.66}

\title[FRB Lensing]{Gravitational lensing in the presence of plasma scattering with application to Fast Radio Bursts}
\author[Kumar \& Beniamini]{
	Pawan Kumar$^1$\thanks{pk@astro.as.utexas.edu} \& Paz Beniamini$^{2,3}$\thanks{pazb@openu.ac.il}\\
	$^1$Department of Astronomy, University of Texas at Austin, Austin, TX 78712, USA\\
	$^{2}$Department of Natural Sciences, The Open University of Israel, P.O Box 808, Ra'anana 4353701, Israel\\
	$^{3}$Astrophysics Research Center of the Open university (ARCO), The Open University of Israel, P.O Box 808, Ra'anana 4353701, Israel}

\begin{document}
	\label{firstpage}
	\pagerange{\pageref{firstpage}--\pageref{lastpage}}
	\maketitle
	
	\begin{abstract}
		We describe how gravitational lensing of fast radio bursts (FRBs) is affected by a plasma screen in the vicinity of the lens or somewhere between the source and the observer. Wave passage through a turbulent medium affects gravitational image magnification, lensing probability (particularly for strong magnification events), and the time delay between images. The magnification is suppressed because of the broadening of the angular size of the source due to scattering by the plasma. The time delay between images is modified as the result of different dispersion measure (DM) along photon trajectories for different images. Each of the image lightcurve is also broadened due to wave scattering so that the images could have distinct temporal profiles. The first two effects are most severe for stellar and sub-stellar mass lens, and the last one (scatter broadening) for lenses and plasma screens at cosmological distances from the source/observer. This could limit the use of FRBs to measure their cosmic abundance. On the other hand, when the time delay between images is large, such that the lightcurve of a transient source has two or more well separated peaks, the different DMs along the wave paths of different images can probe density fluctuations in the IGM on scales $\lesssim 10^{-6}$\,rad and explore the patchy reionization history of the universe using lensed FRBs at high redshifts. 
		Different rotation measure (RM) along two image paths can convert linearly polarized radiation from a source to partial circular polarization.
	\end{abstract}

	
	
	
	
	\section{Introduction}
	\label{sec:intro}
	The probability of strong lensing of a compact source at redshift larger than one, to magnification $>\mu$, by an intervening galaxy is $p(>\mu) \sim 0.3 \Omega_{\rm gal}/\mu^2$ for $z\gtrsim 2$ and $P(>\mu)\approx \Omega_{\rm gal} z^2/(4\mu^2)$ for $z\ll 1$ \citep{NB1996}.
	; where $\Omega_{\rm gal}$ is the mean mass density in galaxies divided by the critical mass density. Strong lensing occurs when a source has multiple images and typically (depending on the mass density in the lens) requires $\mu \gtrsim 1$. For a galaxy of mass $\sim 10^{11}M_{\odot}$ at a Gpc distance from both the source and the observer, the angular separation between the images is of order a few arc-seconds, and the travel time difference of order $\sim 10$ days. For $\mu \gta 1$, the corresponding lensing probability is $\sim 6{\rm x}10^{-3}(\Omega_{11}/0.02)$ (where $\Omega_{11}$ is $\Omega_{\rm gal}$ for galaxies of mass $\sim 10^{11}M_{\odot}$). Thus, in an FRB survey of 10$^4$ sources, we expect of order 60 lenses by intervening galaxies. In addition, considering that the population of FRB-repeaters is about 20\% of the entire FRB  population, about 12 repeating FRBs in the survey should be lensed. The expected number of FRB lenses is small. However, considering that the duration of FRBs is typically a few milli-seconds, one can determine the travel time difference along multiple paths to better than about a ms or about one part in a billion. This remarkable accuracy, which is better than any other class of astronomical object, has led  a number of people e.g., \citep{Eichler2017,Zitrin2018,Li2018,Liu2019,Wucknitz2021,Connor2022,Leung2022}, to suggest that FRB lensing can be a good probe of cosmology. In particular, it has been suggested that this type of lensing delay from FRB-repeaters could be used to measure the Hubble constant \citep{Zitrin2018,Li2018,Liu2019,Wucknitz2021}, by observing the rate of change in the lensing delay (the difference between the arrival of signals from two images of the same burst) over a period of years.
	
	FRBs can also be micro-lensed by a stellar object. The probability for that at $z\lesssim 0.5$ is of order\footnote{This $\Omega_{M_{\odot}}$ estimate is from \citealt{MD2014}.}, $\tau \sim 2.5\times 10^{-4}(\Omega_{M_{\odot}}/0.004)$ for a solar mass object (where $\Omega_{M_{\odot}}$ is the mass density in $\sim 1M_{\odot}$ objects at $z\lesssim 0.5$ divided by the closure density). The lensing probability is significantly larger (by a factor of up to $\sim 50$) if all of dark matter were in $\sim M_{\odot}$ primordial black holes \citep{Carr1974}. Thus, observational limits on $\tau(M)$ can constrain the density of primordial black holes with mass $\sim M$.
	One noteworthy advantage of observing micro-lensing of FRBs by objects with stellar or smaller mass is that it allows to probe the regime of physical, rather than geometrical, optics.
	Gravitational lensing is typically considered in the latter regime.
	This is in large part due to the fact that one needs a source whose size is smaller than $\sim 10^{12}(M/M_{\odot})^{-1/2}(\nu/\mbox{GHz})^{-1}(D/\mbox{Gpc})^{-1/2}\mbox{ cm}$ \footnote{This requirement ensures that the size of the source is less than $d_{\rm SO} \pi \theta_{\rm F}^2/\theta_{\rm E}$ (where $d_{\rm SO}$ is the source-observer distance, and $\theta_{\rm F},\theta_{\rm E}$ are correspondingly the Fresnel and Einstein angular scales) and therefore that the time-delay between the two images is shorter than a wave period.} (\citealt{ND1999}; where $D$ is the distance to the source and we considered for clarity the case in which the lens is halfway between the source and the observer). This condition is not easily satisfied by sources observable at cosmological distances. 
	It has been argued that the coherent nature of FRBs, their detection up to cosmological distances and the potentially small sizes of their sources, make it possible to use them to probe wave effects of plasma and gravitational lensing \citep{GrilloCordes2018,Cordes2017b,KKSX2020,Jow2020}. In the physical optics regime (unlike in geometrical optics) micro-lensing events become frequency dependent. As a result, if the dynamic spectrum of a micro-lensing event can be observed, it allows to uniquely constrain the mass of the lens, $M$, and the combination $\mu_{\rm rel}^2D$ (where $\mu_{\rm rel}$ is the angular velocity of the lens relative to the source in the plane of the sky). This is as opposed to geometrical micro-lensing where only the combination $M/\mu_{\rm rel}^2D$ can be measured.
	
	A different application of micro-lensing of FRBs by stellar mass objects is to constrain the possibility that dark matter is composed of massive compact halo objects (MACHOS) in the range of several to hundreds of solar masses \citep{Munoz2016}. The idea suggested by these authors uses the fact that the time difference between two (unresolved) `images' is a direct measure of the lens mass. Therefore, if this time difference is longer than the duration of an FRB burst, micro-lensing will leave a measurable signature on the lightcurve. This can then be used to constrain the optical depth of lenses with a given mass range. 
	
	As apparent from the above description, many applications of FRB lensing rely on the fact that cosmological FRBs can often be assumed to be effectively point sources. However, the spectra of many FRBs is often seen to suffer from spectral decoherence \citep[e.g.][]{Bannister+19}, and their light-curves show signatures of scatter broadening \citep[e.g.][]{Thornton+13}. These signatures indicate that the FRB wave suffers from scintillation as it passes through a turbulent plasma screen (or multiple screens) on its path from the source to us. Indeed, multi-path propagation induced by plasma scintillation can lead to other notable effects on the observed FRB signals, such as induced temporal variability \citep{BK2020} and depolarization / induced circular polarization \citep{BKN2022}.
	As plasma scintillation is commonly inferred to affect FRB observations, it is natural to ask how it affects gravitational lensing. This is the topic of the present work. As we will show below, perhaps the most important effect of plasma scintillation on lensing is that it effectively increases the size of the source (as mentioned in \citealt{cordes2019}), and therefore suppresses lensing effects for lenses below a critical mass.
	
	The effects of uniform and non-uniform plasma on gravitational lensing by a point source was considered also by \cite{Bisnovatyi-Kogan2010,Bisnovatyi-Kogan2017}. \cite{Eshleman1979} explored how the magnification of a background source's brightness due to gravitational lensing by the Sun is limited by waves traveling through the solar corona. These authors, however, did not consider wave scattering in turbulent plasma, which is the topic of the current work.
	
	The paper is organized as follows. We begin in \S \ref{sec:basic} with a description of some important timescales and length-scales related to gravitational lensing and plasma scintillation. This sets the scene to understand when plasma scintillation affects can not be ignored in the treatment of lensing. In \S \ref{lens-mag-dt} we focus on a simple, but very useful test case of the effects of plasma scattering on lensing from a point source. This case can be worked out in some detail and allows to estimate the effects of a plasma screen on the lensing magnification and on the lens delays as reflected in the FRB lightcurve. In \S \ref{sec:circular}, we describe ways in which gravitational lensing in the presence of plasma scattering can provide unique constraints on cosmology and in addition show that gravitational lensing can enhance the degree of induced circular polarization due to passage of the FRB wave through a magnetized plasma screen. We conclude in \S \ref{sec:conc}. 
	
	\section{Basic considerations}
	\label{sec:basic}
	The travel time difference along two different paths corresponding to two gravitational lens images of an FRB by a point lens of mass $M_{\rm l}$ is approximately $R_{\rm s}/c=2G M_{\rm l}/c^3$; where $R_{\rm s}\equiv 2 G M_{\rm l}/c^2$ is the lens' gravitational radius, and $c$ is the speed of light in vacuum. For a stellar mass lens, this time difference is $\Delta t_g\sim 10\mu$s. Coherent FRB radio waves traveling along two different trajectories could interfere and physical optics effects could influence the magnification of the observed flux when the lens mass is small, e.g. \cite{Jow2020}.
	
	In addition, an EM wave propagating through a medium of non-zero electron density moves at a speed slightly smaller than $c$ and that contributes to a delay in photon arrival time that is different for different trajectories. We show below that this plasma effect cannot be ignored for stellar mass lens. The dispersion relation and group speed for EM waves in plasma are
	\begin{equation}
		\omega^2 = \omega_{\rm p}^2 + c^2 k^2, \quad {\rm and} \quad v_{\rm g} = \frac{d\omega}{dk} \approx c\left[ 1 - \frac{\omega_{\rm p}^2}{2\omega^2}\right]
	\end{equation}
	where the plasma frequency is
	\begin{equation}
		\omega_{\rm p} = \left(\frac{4\pi q^2 n_{\rm e}}{m}\right)^{1/2}
	\end{equation}
	and $n_{\rm e}$, $q$, and $m$ are electron density, charge and mass respectively. Thus, the delay in photon arrival when it travels a distance $d_s$ through plasma is
	\begin{equation}
		t_{\rm p} = \int_0^{d_s} \frac{dr}{v_{\rm g}} - \frac{d_s}{c} = \int_0^{d_s} \frac{dr}{c} \frac{2\pi q^2 n_{\rm e}}{m \omega^2} = {\rm 4.2\, ms}\, \nu_9^{-2}\, {\rm DM},
		\label{dtp}
	\end{equation}
	where
	\begin{equation}
		{\rm DM} \equiv \int_0^{d_s} \frac{dr}{1\, {\rm pc}}\, n_{\rm e}
	\end{equation}
	is the dispersion measure along the photons trajectory, with $n_{\rm e}$ measured in CGS units, and 1 pc = 3.14x10$^{18}$ cm. 
	
	Gravitational images in the lens plane are separated by a distance of order the Einstein radius which is defined by
	\begin{equation}
		R_{\rm E} = \left( \frac{2 R_{\rm s} d_{\rm SL} d_{\rm LO}}{d_{\rm SO} }\right)^{1/2} ,
		\label{RE}
	\end{equation}
	where $d_{\rm SL}$ is the distance between the lens and source, $d_{\rm LO}$ is distance between the lens and observer, and $d_{\rm SO}$ is the distance between the source and the observer. It is useful to define also the Einstein angle
	\begin{equation}
		\theta_{\rm E}={R_E \over d_{\rm LO}} = \left(\frac{2 R_{\rm s} d_{\rm SL}}{d_{\rm SO} d_{\rm LO}}\right)^{1/2} ,
		\label{thetaE}
	\end{equation}
	
	For a stellar mass lens at a cosmological distance, $R_{\rm E}\sim 10^{17}$cm. Even if the difference in DM along the two photon trajectories is only $\delta \mbox{DM}=10^{-3}$ cm$^{-3}$ pc -- due to fluctuations in the electron density on a scale $R_{\rm E}$ -- the travel time difference along the two trajectories due to plasma effects is $\Delta t_p=4\mu$s at 1 GHz (eq. \ref{dtp}), i.e. of the same order as the gravitational time delay. Density fluctuations of this magnitude are present when the outer scale of turbulence for the IGM is of order $10^{22}$cm or less and this is discussed further in \S\ref{time-delay}. Furthermore, even if $\Delta t_p\ll \Delta t_g$, plasma scattering may significantly modify lensing. Indeed, it turns out that typically the limiting condition for modification of gravitational lensing by plasma effects is related to the ratio of the plasma scattering angle and $\theta_{\rm E}$ (\S \ref{sec:magnification}).
	
	For a lens of mass larger than $\sim$10$^2$ M$_\odot$, the time delay between different images is $\Delta t_g\gta$1ms. In this case (barring plasma effects) no interference between different lensed images is expected for most FRBs as their durations are a few ms. The magnification of each image is set by interference of the bundle of rays, the {\bf Fresnel bundle}, that have traveled along different trajectories with travel-times that lie within a few wave periods of each other. The width of the {\bf Fresnel bundle}, $R^2_F \sim \lambda\min\{d_{\rm SL}, d_{\rm LO}\}$; $\lambda$ is photon wavelength. The deflection of rays across the Fresnel bundle does not change much as long as $R_F \ll R_{\rm E}$.
	And that means that geometrical optics is a good description of gravitational lensing as long as $R_{\rm s}\gg \lambda$ and $\theta_{\rm S} \not\ll \theta_E$; where $\theta_{\rm S}$ is the angle between the observer-source and observer-lens lines.
	
	This can also be understood by considering the time delay between the images.
	For a point mass lens, and a source angle position $\theta_{\rm S}\lesssim \theta_E$, the time delay between the two images is $\Delta t\sim (4 R_{\rm s}/c)(\theta_{\rm S}/\theta_E)$ (see \S \ref{time-delay} for the derivation of this result; eq. \ref{delay1}). The delay is less than the wave period, or the images interfere, when $\theta_{\rm S}<\theta_{\rm F}^2/\theta_{\rm E}$; where the Fresnel angle is defined to be \begin{equation}
		\theta_F = \left[ \frac{\lambda d_{\rm SL}}{d_{\rm LO} d_{\rm SO}}\right]^{1/2}.
	\end{equation} 
	Since, the most likely lensing event is one where $\theta_{\rm S}\sim \theta_{\rm E}$, we see that typically interference is important when $\theta_{\rm F}<\theta_{\rm E}$ or equivalently $R_{\rm s} < \lambda$. Therefore, geometrical optics is a good description of gravitational lensing for most cases of interest\footnote{
		The interference between bundles of rays associated with the two images can modify the overall magnification as given by the geometrical optics even when $R_{\rm s} \gg\lambda$. This can occur if the difference in travel time between images is smaller than the coherence time, $T_c$, of the transient source being lensed. This translates to the condition that $R_{\rm s} \lta c T_c$ for physical optics to be important. However, it should be noted that the interference between the two images can increase the total magnification by at most a factor two, but destructive interference can reduce the total flux for the two images to almost zero.}. However, plasma effects can be important more widely as discussed above, and that is the subject of this paper.
	
	The effect of plasma on image magnification is important when the scattering angle for waves traveling through the plasma screen ($\theta_{\rm scat}$) is of order $\theta_{\rm S}$. The scattering broadens the angular size of the source to $\theta_{\rm scat}$, and limits the magnification to $\theta_{\rm E}/(2\theta_{\rm scat})$ as we show in \S\ref{lens-mag-dt}.
	
	\section{Gravitational lensing due to a point mass in the presence of plasma scattering screen}
	\label{lens-mag-dt}
	
	The flux observed from an astronomical source when photons travel through a gravitational potential and plasma on their way to the observer is given by
	\begin{equation}
		f(\omega, {\vec\theta_{\rm s}}) =  \frac{1}{i\theta_F^2} \int d^2\theta \, \exp\left\{ \frac{i\pi |\vec\theta - \vec{\theta}_s|^2}{\theta_F^2} - i\omega \left[ \psi(\vec\theta) - \delta t_p(\vec\theta)\right]\right\} 
		\label{path-integral}
	\end{equation}
	where $\vec\theta_s$ is the angular location of the source wrt observer-lens line, and $\vec\theta$ is angular position of a point in the lens plane again wrt observer-lens line. The first term in the bracket, $\pi|\vec\theta - \vec\theta_s|^2/\theta_F^2$ is the geometric phase shift suffered by the wave as a result of its trajectory not being a straight line from the source to the observer,
	\begin{equation}
		\psi(\vec\theta) = \int d\ell\, \frac{\Phi(\vec x)}{c^3} (1 + n_r^2)
	\end{equation}
	is the time delay due to travel in the gravitational potential $\Phi$\footnote{General relativistic effects have been neglected in the expression for $\psi$}, where the integration is done over the photon path, $n_r$ is the radial component of unit vector that is tangent to the photon path, and $\delta t_p$ is the time delay suffered by the wave due to propagation through plasma. The expression for $\psi$ is valid only for spherically symmetric gravitational potential, and the radial component is wrt the center of the symmetry. This equation is a generalization of the result given in \citep{ND1999} to include plasma effects. For a point lens of mass $M$, as per \cite{Schneider1992}
	\begin{equation}
		\psi(\vec\theta) = \frac{4 GM}{c^3} \log|\vec\theta| = \frac{2 R_{\rm s}}{c} \log|\vec\theta|.
		\label{gravity-delay}
	\end{equation}
	
	Much of the contribution to the integral in eq. (\ref{path-integral}) comes from the extrema of the phase function in the exponent. In the absence of the turbulent plasma screen, the phase has two extrema for a point lens corresponding to two gravitational lens images. However, for waves passing through a turbulent plasma screen $\delta t_p(\theta)\propto \theta^{5/6}$ and it fluctuates on an eddy length scale of $\ell_{\phi}$. Thus, the exponent has large number of extrema that correspond to waves being scattered from different segments of the plasma screen to arrive at the observer. The area around each of these extrema that contributes to the integral is of angular size $\delta\theta_{\phi}\sim \ell_{\phi}/d_{\rm LO}$ such that the exponent changes by $\sim \pi$ radian across the area. Expanding the exponent about one of the image locations in the absence of the plasma, $\theta_{\rm I}$, we see that when $\delta \theta \gta \theta_F^2/\delta\theta_{\phi}$, the phase is dominated by the geometrical path length term and grows as $\delta\theta^2$, hence the integrand oscillates rapidly and does not contribute much to the integral. This can be cast in the usual picture that the radius of the screen from which photons arrive at the observer is $\sim (\theta_F/\delta\theta_{\phi})^2\ell_{\phi}$, which is defined as the refractive scale. Waves are scattered by different patches of size $\ell_{\phi}$ within the refractive scale and add incoherently at the observer location. 
	
	In the next sub-section we calculate the effect on magnification of gravitationally lensed images due to the turbulent plasma screen. And in \S\ref{time-delay} we discuss how the time delay between different lensed images are modified by the plasma screen.
	
	\subsection{Effect of plasma screen on lens magnification}
	\label{sec:magnification}
	\begin{figure}
		\centering
		\includegraphics[width = 0.4\textwidth]{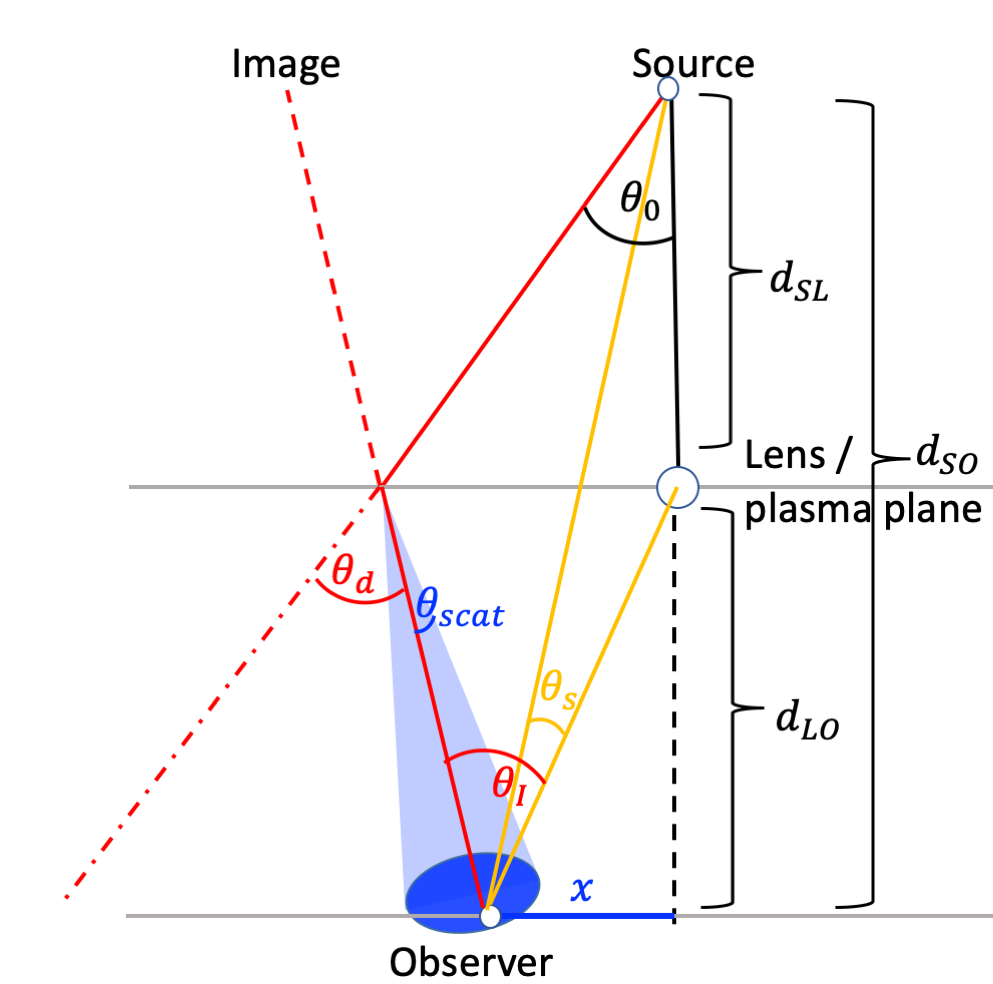}
		\caption{ Schematic figure representing the geometry of the source - lens - plasma - observer system. The plasma plane is assumed here to coincide with the lens plane.}
		\label{fig:schematic}
	\end{figure}
	
	We refer to Fig. \ref{fig:schematic} for the derivation of magnification of gravitationally lensed point-source in the presence of a scattering plasma screen. The derivation assumes that the plasma lies in the lens plane to simplify the algebra. However, the results are broadly applicable when the lens and plasma planes are separate except for some geometrical factors which can be important when the distance between the two planes becomes comparable to the smaller of the distances between the source and the lens or lens and the observer (the relevant modification in this case is described in \S \ref{sec:conc}).
	
	The lens equation is
	\begin{equation}
		\theta_{\rm I} d_{\rm SO} = \theta_d d_{\rm SL} + \theta_{\rm S} d_{\rm SO},
	\end{equation}
	where the deflection angle 
	\begin{equation}
		\theta_d = \frac{2 R_{\rm s}}{\theta_0 d_{\rm SL}} = \frac{2 R_{\rm s}}{\theta_{\rm I} d_{\rm LO}},
	\end{equation}
	$R_{\rm s} = 2 GM_{\rm l}/c^2$ is the gravitational radius of the lens. Substitution of $\theta_d$ in the lens equation gives the standard result for image location in terms of the angular position of the source wrt the lens:
	\begin{equation}
		\label{eq:thI}
		\theta_{\rm I} = \frac{1}{2} \left[ \theta_{\rm S} \pm \left( \theta_{\rm S}^2 + 4\theta_{\rm E}^2\right)^{1/2}\right].
	\end{equation}

	We calculate the magnification of the image that is on the same side of the lens-source axis as the observer. The magnification of the other image can be worked out in an equivalent way by considering the image with the minus side in eq. \ref{eq:thI}.
	One way to calculate the magnification by the lens, which can be easily generalized to include the plasma scintillation, is to consider a bundle of rays between two cones of angles $\theta_0$ and $\theta_0 + \delta\theta_0$, with the apex of these cones at the source. The rays are bent by the lens so that the cross-sectional area of this bundle in the observer plane is $2\pi x\delta x$ instead of $2\pi \theta_0 \delta\theta_0 d_{\rm SO}^2$ as would occur with no lensing. $x,\delta x$ are given by 
	\begin{equation}
		x = \theta_0 (1 - \theta_{\rm E}^2/\theta_{\rm I}^2) d_{\rm SO}\quad \& \quad
		\delta x = d_{\rm SO} \delta\theta_0 \left[ 1 + \theta_{\rm E}^2/\theta_{\rm I}^2\right].
		\label{theta-x}
	\end{equation}
	The magnification in the absence of the scattering screen is $\mu_{gl}=\theta_0\delta\theta_0 d_{\rm SO}^2/(x\delta x)$, which reduces to the standard expression, i.e.
	\begin{equation}
		\mu_{gl} = \frac{1}{1 - \theta_{\rm E}^4/\theta_{\rm I}^4}.
		\label{Mgl}
	\end{equation}
	
	The effect of the plasma screen is that a bundle of rays within a finite angle between $\theta_{-}$ \& $\theta_{+}$ are scattered and their intersection with the observer plane has a larger area than it would in the absence of scattering. We make use of the circular symmetry about the source--lens axis to simplify the calculation of magnification.
	
	The magnification is calculated approximately by making the simplifying assumption that a bundle of rays from the point source with cross-section of $\ell_{\phi}$ in the lens plane are scattered by the column of turbulent eddies in the plasma screen into a cone of angle $\theta_{\rm scat}$, and these rays are bent by an angle $\theta_d$ by the point lens as they travel toward the observer. The scattered rays in the bundle lie between distances $x(\theta_0)\pm \theta_{\rm scat} d_{\rm LO}$ from the source-lens line in the observer plane (Fig. \ref{fig:schematic} shows the geometry). If the source is at an angle $\theta_{\rm S}$ and the observer is located at distance $x_{\rm o}$ wrt source-lens line (see Fig. \ref{fig:schematic}), then the observer will lie in the scattering cone when the outer edge of the cone lies between $x_{\rm o}\pm 2 \theta_{\rm scat} d_{\rm LO}$. The source angle $\theta_{\rm S}$ and $x_{\rm o}$ are related by (as the geometry in Fig. \ref{fig:schematic} makes clear)
	\begin{equation}
		x_{\rm o} = \frac{\theta_{\rm S} d_{\rm SO} d_{\rm LO}}{d_{\rm SL}}.
		\label{xo}
	\end{equation}
	Thus, the range of angles $\theta_0$, $[\theta_-,\theta_+]$, from which rays arrive at the observer after crossing the lens and plasma screen is given by
	\begin{equation}
		\theta_\pm d_{\rm SO} \left[ 1 - {\theta_{\rm E}^2 d_{\rm LO}^2\over \theta_\pm^2 d_{\rm SL}^2} \right] = \max\Big\{0, x_{\rm o} \pm \theta_{\rm scat} d_{\rm LO} \Big\}. 
		\label{thetapmA}
	\end{equation}
	Where we made use of equation \ref{theta-x} that expresses the distance from the source-lens line that a ray that started out from the source at angle $\theta_\pm$ will be at in the observer plane in absence of the scattering plasma.
	
	Thus, the picture is that a conical bundle of rays that leave the source between angles $\theta_-$ and $\theta_+$ arrive at the observer after crossing the lens-plasma plane. And this bundle of rays occupy a circular annulus in the observer plane of radii
	\begin{equation}
		x_{\pm} = \max\Big\{0, x_{\rm o} \pm 2\theta_{\rm scat} d_{\rm LO}\Big\}.
	\end{equation}
	with the observer lying in the middle of this ring. Since the energy in a bundle of rays between the cones of angle $\theta_\pm$ flows through the observer plane between $x_\pm$, the average magnification at the observer location as a result of gravitational lensing and scattering by turbulent plasma is given approximately by
	\begin{equation}
		\mu \approx { d_{\rm SO}^2\left[ \theta_+^2 - \theta_-^2\right] \over \left( x_+^2 - x_-^2\right)/2 },
		\label{M1}
	\end{equation}
	The factor 2 in the denominator is geometric in origin and accounts for the fact that the flux in the middle of a ring formed by superposition of cones rotated about the source-lens axis is larger than the average by a factor $\sim 2$.
	
	Equation (\ref{thetapmA}) for $\theta_\pm$ can be rewritten using eq. \ref{xo} as
	\begin{equation}
		{d_{\rm SL} \theta_\pm \over d_{\rm LO}} \left[ 1 - {\theta_{\rm E}^2 d_{\rm LO}^2\over \theta_\pm^2 d_{\rm SL}^2} \right] = \max\Big\{0, \theta_{\rm S} \pm {\theta_{\rm scat} d_{\rm SL}\over d_{\rm SO}} \Big\}.
		\label{thetapmB}
	\end{equation}
	This equation is further simplified when expressed in terms of 
	\begin{equation}
		\label{eq:thscatp}
		\theta^I_\pm \equiv \theta_\pm \left[{d_{\rm SL}\over d_{\rm LO}}\right] \quad \& \quad \theta'_{\rm scat} \equiv \theta_{\rm scat} \left[ {d_{\rm SL}\over d_{\rm SO}} \right]
	\end{equation}
	and it takes the following form
	\begin{equation}
		\theta^I_\pm \left[ 1 - {\theta_{\rm E}^2 \over {\theta_\pm^I}^2} \right] = \max\Big\{0, \theta_{\rm S} \pm \theta'_{\rm scat} \Big\}.
		\label{theta-pm}
	\end{equation}
	This is the standard point mass lens equation with source located at $\max\Big\{0, \theta_{\rm S} \pm \theta'_{\rm scat} \Big\}$. The physical interpretation of the result is straightforward, viz. the effect of a scattering screen is to broaden a point source so that it has an effective angular size of $\theta'_{\rm scat}$.
	
	The equation for magnification (\ref{M1}) simplifies to
	\begin{equation}
		\mu \approx { 2\left[ {\theta^I_+}^2 - {\theta^I_-}^2\right] \over \left(\theta_{\rm S} + 2\theta'_{\rm scat}\right)^2 - \left(\max\big\{0, \theta_{\rm S} - 2\theta'_{\rm scat}\big\}\right)^2}.
	\end{equation}
	
	It is easy to see that this equation reduces to the standard form, i.e. eq. \ref{Mgl}, in the absence of the plasma ($\theta_{\rm scat}=0$); the numerator in that limit is $2(d\theta_{\rm I}^2/d\theta_{\rm S})\delta\theta=4(\delta\theta)\theta_{\rm I}^3/(\theta_{\rm I}^2 + \theta_{\rm S}^2)$, where $\delta\theta=2\theta_{\rm scat}'$.
	We also note that in the limit $\theta_{\rm scat}'\gg\theta_{\rm E},\theta_{\rm s}$, eq. \ref{Mgl} reduces to $\mu \to 1/2$. This might appear counter-intuitive as in the presence of strong scattering we would expect no magnification or $\mu\approx 1$. However, in this limit, the area in the lens plane that is traversed by rays that compose image 1 (on the same side of the lens-source axis as the observer) almost completely overlaps the area in the lens plane of the rays forming image 2 (on the opposite side of the said axis). This means that the images can no longer be separated in any physical way (i.e. either in terms of their angular position on the sky or in terms of the time delays associated with them), and one can no longer treat the two images separately. The magnification for the two images together approaches unity in this case as one would expect.
	
	For the case where $\theta_{\rm E} > \theta_{\rm scat}' \gta \theta_{\rm S}$, $\theta_- = \theta_{\rm E}$ as per equation (\ref{theta-pm}), and the magnification is
	\begin{equation}
		\mu = {(\theta_{\rm S} + \theta_{\rm scat}') \over (\theta_{\rm S} + 2\theta_{\rm scat}')^2 } \left[ \theta_{\rm S} \pm \sqrt{\theta_{\rm S}^2 + 4\theta_{\rm E}^2} \right].
	\end{equation}
	The presence of the scattering screen decreases the magnification by a factor $\sim 2\theta_{\rm scat}'/\theta_{\rm S}$ in this case.
	
	The magnification in the opposite case of $\theta_{\rm scat}' < \theta_{\rm S}$ is
	\begin{equation}
		\mu = { (\theta_{\rm S} + \theta_{\rm scat}') \theta^I_+ - (\theta_{\rm S} - \theta_{\rm scat}') \theta^I_- \over 4\theta_{\rm S} \theta_{\rm scat}'}.
	\end{equation}
	
\begin{figure}
		\centering
		\includegraphics[width = 0.48\textwidth]{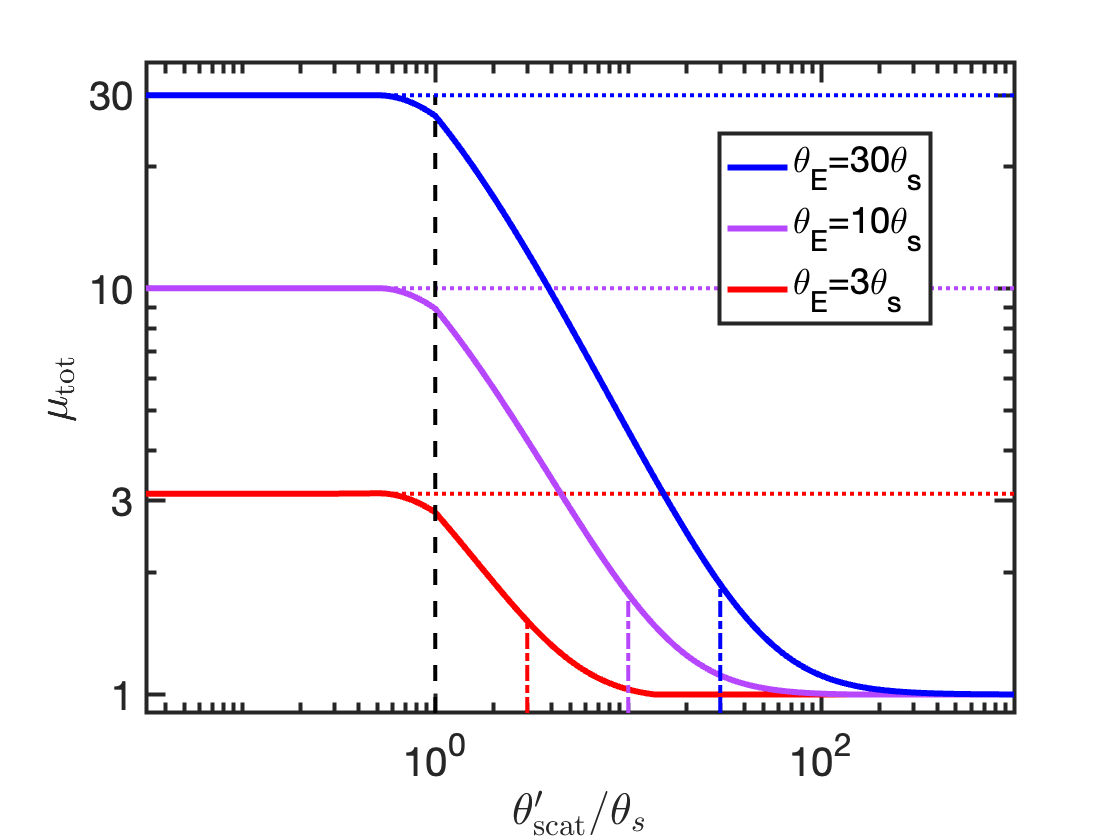}
		\caption{ Effect of plasma scattering on gravitational lensing magnification. Solid lines depict the total magnification (i.e. the sum of the two images) as a function of $\theta_{\rm scat}'/\theta_{\rm s}$ and for different values of $\theta_{\rm E}/\theta_{\rm s}$ (shown by different colors). Horizontal dotted lines show the values of the magnification for the same parameter but without plasma scattering. A dashed (dot dashed) vertical line represents $\theta_{\rm scat}'=\theta_{\rm s}$ ($\theta_{\rm scat}'=\theta_{\rm E}$). Plasma scattering causes lensing magnification to be suppressed by a factor $\sim 2\theta_{\rm scat}'/\theta_{\rm s}$ for $\theta_{\rm scat}'>\theta_{\rm s}$ until there is virtually no magnification when $\theta_{\rm scat}'\approx \theta_{\rm E}$.}
		\label{fig:Mscatter}
	\end{figure}

	The total magnifications (summing over the two images) for a few values of $\theta_{\rm S}/\theta_{\rm scat}'$ are shown in Fig. \ref{fig:Mscatter}.
	The magnification in the presence of a scattering screen for the sum of the flux for the two images is capped at $\theta_{\rm E}/2\theta'_{\rm scat}$ (see. Fig. \ref{fig:Mscatter}). Thus, there is a minimum lens mass, $M_{\rm min,\mu}(\mu_{\rm max})$, below which the magnification cannot be larger than $\mu_{\rm max}$. This mass is given by
	\begin{equation}
		\label{eq:Mlimmag}
		\begin{aligned}
			M_{\rm min,\mu}(\mu_{\rm max}) & \sim {\mu_{\rm max}^2 \theta_{\rm scat}^2 c^2\over G} {d_{\rm SL} d_{\rm LO}\over d_{\rm SO}} \\ & \sim (7\, M_\odot) \mu_{\rm max,1}^2 \theta_{\rm scat,-9}^2 d_{\rm LO,22} (d_{\rm SL}/d_{\rm SO}).
		\end{aligned}
	\end{equation}
	
	The scatter broadening timescale for a lensed burst scales as\footnote{The reason that the scatter broadening time scales as $\theta'^2_{\rm scat}$ and not as $\theta_E \theta'_{\rm scat}$ is that at the extrema points of the path-integral in eq. \ref{path-integral}, the first derivative of the geometrical + gravitational contributions to the phase vanishes. In other words, their respective contributions cancel each other to first order, and for small deviations around $\theta_{\rm I}$, only the second order term contributes to the time delay.} $\theta'^2_{\rm scat}$, and is given by
	\begin{equation}
		\label{eq:tsc}
		t_{\rm sc}\approx \frac{(\theta_{\rm scat}'d_{\rm LO})^2}{c}\frac{d_{\rm SO}}{d_{\rm LO} d_{\rm SL}}=\frac{\theta_{\rm scat}^2 d_{\rm SL} d_{\rm LO}}{cd_{\rm SO}}=\frac{2R_{\rm s}}{c} \left(\frac{\theta_{\rm scat}'}{\theta_{\rm E}}\right)^2
	\end{equation}
	As will be shown in \S \ref{time-delay}, the first term on the R.H.S. is of order the gravitational+geometrical time delay between the images, $\Delta t_{g}$. For $\theta_{\rm scat}'\ll \theta_{\rm E}$, we see from eq. \ref{eq:tsc} that $t_{\rm sc}\ll \Delta t_g$. This timescale is also shorter than $t_{\rm FRB}$, if
	\begin{equation}
		\label{eq:dsc}
		\min(d_{\rm LO},d_{\rm SL})<d_{\rm sc}\equiv (3\times 10^{25} {\rm cm})\, t_{\rm FRB,-3}\theta_{\rm scat,-9}^{-2}.
	\end{equation}
	
	For lens mass below $M_{\rm min,\mu}(\mu_{\rm max}=1)$, the temporal broadening for each of the `image' pulse due to plasma turbulence is larger than the delay between their arrival times ($\Delta t_{g}$). As a result, the presence of gravitational lensing would be very hard (if not impossible) to infer from the lightcurve below this mass.
	
	Even if the scatter broadening timescale is short compared to the time delay between the images, it might still be larger than the intrinsic duration of the FRB. In such a situation, a delay between the image arrival times would be detectable, but the shape of the lightcurves for the two images would be different. It is important to note that lensing could still be identifiable in this case for non-repeating FRBs, as the later image lightcurve will have a very similar DM and likely an indistinguishable angular position relative to that of the first image. The situation is more complicated for a repeater, as the different looking lightcurves of different images could be confused with different bursts from the source. Lensing could still be identified by consistent delays between pairs of bursts (for a point lens) from the source, or perhaps by sophisticated data analysis that can deconvolve the different scatter broadening of the two images. 
	
	The scattering angle $\theta_{\rm scat}$ in eqns. \ref{eq:Mlimmag}, \ref{eq:dsc} can be estimated as $\theta_{\rm scat}\sim \lambda/\ell_\phi$, where $\ell_\phi$ is the diffraction scale for the turbulent plasma screen, and $\lambda$ is the wavelength of FRB radiation. We take a brief detour here to provide an estimate for $\ell_\phi$, and thus $\theta_{\rm scat}$. 
	
	Let us consider that the electron density fluctuation in the turbulent medium is a power-law function in the inertial sub-range between length scale $\ell_{\rm min}$ and $\ell_{\rm max}$, and is given by
	\begin{equation}
		\delta n_{\rm e}(\ell) = n_{\rm e}(\ell/\ell_{\rm max})^\alpha.
	\end{equation}
	The index $\alpha = 1/3$ is for Kolmogoroff density fluctuations. The largest eddy size, $\ell_{\rm max}$, is the scale at which energy is injected to maintain the turbulence, and the smallest scale $\ell_{\rm min}$ is determined by dissipation physics of turbulence. The fluctuation in the dispersion measure (DM) for waves traveling through a plasma screen of width ${\rm L}$, at two points separated by length scale $\ell$, is given by
	\begin{equation}
		\delta {\rm DM}(\ell) \sim \delta n_{\rm e}(\ell) (\ell/L)^{1/2} \sim {\rm DM_L}\, (\ell/\ell_{\rm max})^\alpha \, (\ell/\rm{L})^{1/2},
		\label{dDM}
	\end{equation}
	where ${\rm DM_L} = n_{\rm e} {\rm L/(1 pc)}$.
	
	The diffraction scale ($\ell_\phi$) is defined to be the transverse length in the plasma screen such that a wave suffers a differential phase shift of $\sim \pi$ across $\ell_\phi$ after crossing the screen, and is given by
	\begin{equation}
		\ell_{\phi} \sim \left( {m c^2\over q^2 n_{\rm e}\lambda}\right)^{{6\over5}} \ell_{\rm max}^{{2\over 5}} L^{-{3\over 5}}
		\sim (2{\rm x}10^{13} {\rm cm})\, n_{\rm e}^{-{6\over5}} L^{-{1\over 5}} \nu_9^{6\over 5} \left( {\ell_{\rm max}\over L}\right)^{2\over 5}.
		\label{lpi}
	\end{equation}
	The various exponents in the above expression for $\ell_{\phi}$ are for the particular case of $\alpha=1/3$ or the Kolmogoroff spectrum. Strong scattering, or diffractive scintillation, occurs when $\ell_{\phi} < R_F\equiv \theta_F d_{\rm LO}$. 
	
	The scattering angle is approximately equal to the diffraction angle corresponding to size $\ell_\phi$, and can be written using equation \ref{lpi} as
	\begin{equation}
		\theta_{\rm scat} \sim {\lambda\over \ell_\phi} = (1.5{\rm x}10^{-12} {\rm rad})\, n_{\rm e}^{{6\over5}} L^{{1\over 5}} \nu_9^{-{6\over 5}} \left( {\ell_{\rm max}\over L}\right)^{-{2\over 5}}.
		\label{theta-scat3}
	\end{equation}
	
	As an example, a plasma screen at 1 pc of the FRB source with $n_{\rm e}\sim 1$ cm$^{-3}$ and $\ell_{\rm max}/L\sim 10^{-4}$, has $\ell_{\phi}\sim 10^8 \nu_9^{6/5}$ cm. The apparent source size due to scattering by the plasma screen is $d_{\rm SP} \lambda/\ell_{\phi} \sim 10^{11}$cm whereas the Einstein radius for a stellar mass lens within 100 kpc of the source is, $R_{\rm E} \sim 10^{15}$cm. Thus, the plasma screen within a few pc of the source has little effect on the magnification by a stellar mass lens 100 kpc away.
	
	We can now recast the expression for $M_{\rm min,\mu}$ by making use of equation (\ref{theta-scat3}) for $\theta_{\rm scat}$:
	\begin{equation}
		\label{eq:Mlimmag2}
		M_{\rm min,\mu}
		\sim 2{\rm x}10^4 M_\odot\, { \mu_{\rm max,1}^2 {\rm DM}_2^{12\over5} d_{\rm LO,22}\, d_{\rm SL} \over \nu_9^{12\over5} L_{22}^{6\over5} \ell_{\rm max,17}^{4\over5}\, d_{\rm SO}}.
	\end{equation}
	Similarly, we recast eq. \ref{eq:tsc} as
	\begin{equation}
		\label{eq:dsc2}
		\min(d_{\rm LO},d_{\rm SL})<d_{\rm sc}\equiv (10^{26}\, {\rm cm})\, \frac{t_{\rm FRB,-3}\mbox{DM}_2^{12/5}}{L_{22}^{6/5}\nu_9^{12/5} \ell_{\rm max,17}^{4/5}}.
	\end{equation}
	
	\subsubsection{Interference between multiple images due to a lens}
	
	Photons from multiple images of a coherent source would interfere provided that the time delay between images is less than the duration of the burst or the source coherence time, whichever is smaller. Let us assume that there are $N$ images and that the travel time difference wrt to the first image (selected arbitrarily) is $t_i^{l}$ ($t_1^{l}=0$). If there is a plasma screen somewhere between the source and the observer then that would cause an additional time delay for these $N$ images, and affect their interference and the magnification for the combined $N$ images. Let us take the transverse separation between the $i$-th and the first image path in the plasma screen to be $\ell_{\perp i}$. Combining equations (\ref{dtp}) \& (\ref{dDM}) we find the time delay due to plasma for image $i$ to be
	\begin{equation}
		t_i^p \approx {\rm 4.2\, ms}\, \nu_9^{-2}\, {\rm DM}_{\rm L} \left({\ell_{\perp i}\over L}\right)^{1/2} \left({\ell_{\perp i}\over\ell_{\rm max}}\right)^\alpha,
	\end{equation}
	where DM is the mean DM for the plasma screen. If the flux at the observer for image $i$ is $f_i$ then in the Eikonal approximation, the combined flux for the $N$ images is
	\begin{equation}
		f = 2\sum_{i,j} (f_i f_j)^{1/2} \cos\left[\omega(t_i^l + t_i^p - t_j^l - t_j^p)\right].
	\end{equation}
	The flux $f_i$ is also modified by the presence of the plasma screen when $\ell_\phi$ is smaller than the Fresnel length, i.e. in the strong scattering regime. The phase difference between waves along the different image trajectories through the turbulent medium is random because the travel time difference, $t_i^p - t_j^p$, is random for most astrophysical systems of interest as discussed in \S \ref{time-delay}.
	
	\subsection{Optical depth for lensing in the presence of plasma scattering}
	\label{sec:tau}
	We showed in \S \ref{sec:magnification} that the magnification is capped at $\min(\theta_{\rm E}/2\theta_{\rm S},\theta_{\rm E}/2\theta_{\rm scat}')$ in the limit of strong lensing. Therefore, the optical depth for lensing is modified in the presence of plasma.
	
	Consider a distribution of lenses, all with mass $M$ and with an optical depth (uncorrected for scattering) $\tau(>\mu)=\tau_0 \mu^{-2}$ \footnote{$\tau_0$ is directly proportional to the mass density of the lenses, see \S \ref{sec:intro}}. For clarity, we focus here on the case in which the sources under consideration have $z\ll1$, so that cosmological redshift corrections can be neglected, and in which the plasma screen location is in the vicinity of the gravitational lens. We denote a dimensionless distance $x\equiv d_{\rm LO}/d_{\rm SO}$ such that $0<x<1$. 
	For a homogeneous distribution of lenses between the source and observer, the fraction of lenses per dimensionless distance $x$ is $dP_{\rm x}/dx=3x^2$.
	In the absence of plasma scattering, a magnification $\mu>1$ is obtained for sources with $\theta_{\rm s}$ such that 
	\begin{equation}
		\label{eq:thsmu}
		\theta_{\rm s}<\theta_{\rm s,\mu}\equiv \frac{\theta_{\rm E}}{2\mu}=\frac{\theta_{\rm E,0}(1-x)^{1/2}}{2\mu x^{1/2}}
	\end{equation}
	where $\theta_{\rm E,0}\equiv \theta_{\rm E}(x=1/2)$. Assuming an isotropic population of sources and lenses, the fraction of sources with an angle $<\theta_{\rm s}$ is $P_{\rm s}(\theta_{\rm s})=(1-\cos\theta_{\rm s})/2$ or $P_{\rm s}(\theta_{\rm s})=(1/4)\theta_{\rm s}^2$ for $\theta_{\rm s}\ll 1$.
	In the presence of plasma, to obtain magnification $>\mu$, the scattering angle of rays going through the plasma plane must satisfy
	\begin{equation}
		\label{eq:thscatmu}
		\theta_{\rm scat}'<\frac{\theta_{\rm E}}{4\mu} \to \theta_{\rm scat}<\theta_{\rm scat,\mu}\equiv \frac{\theta_{\rm E,0}}{(1-x)^{1/2}x^{1/2} 4\mu}
	\end{equation} 
	where we have used eq. \ref{eq:thscatp} to write $\theta_{\rm scat}'=\theta_{\rm scat}(1-x)$. The distribution of scattering angles for plasma screens depends on the parameters of their turbulence. For generality, we take that to be an unspecified function $P_{\rm scat}(<\theta_{\rm scat})$. We explore specific forms of $P_{\rm scat}$ below.
	
\begin{figure}
		\centering
		\includegraphics[width = 0.48\textwidth]{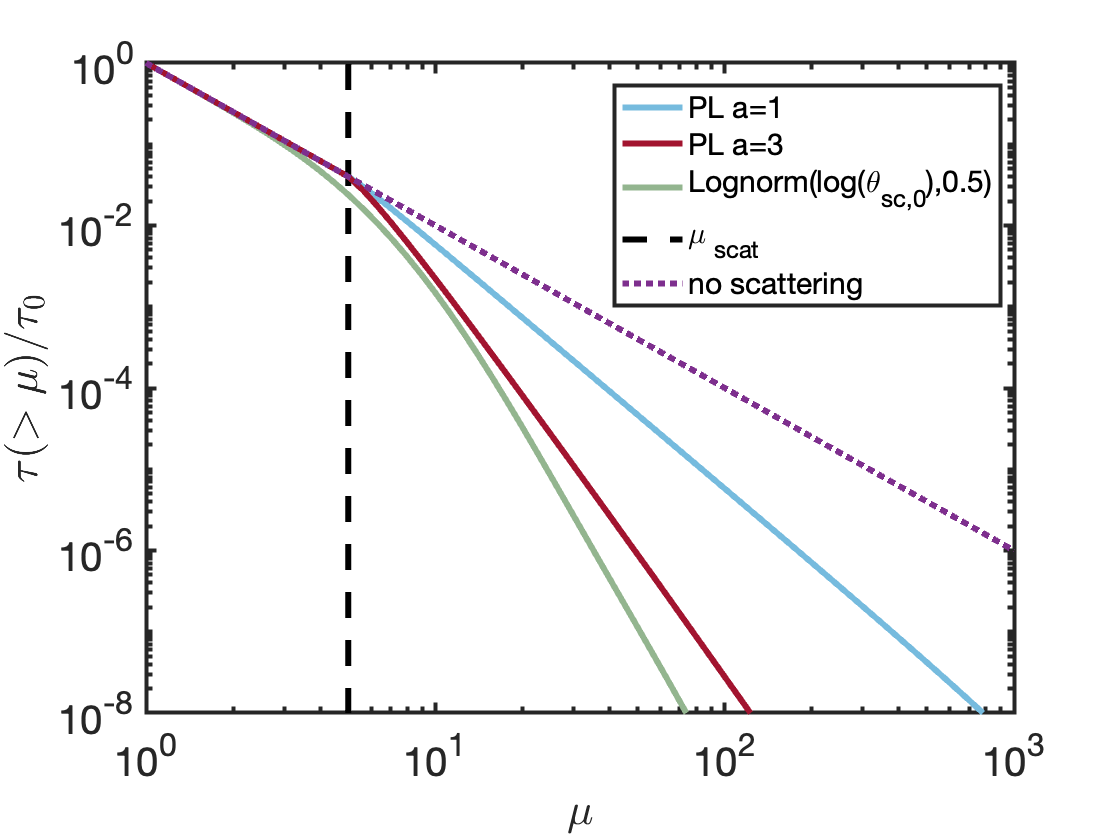}
		\caption{Optical depth for lensing in the presence of plasma scattering (assumed here to take place in the lens plane). Plasma scattering leads to a suppression of the optical depth for magnifications $\mu \gtrsim \mu_{\rm scat}=\theta_{\rm E,0}/2\theta_{\rm scat,0}$. For $\mu\gg \mu_{\rm scat}$, the suppression is by a factor of up to $(\mu_{\rm scat}/\mu)^4$. Results are shown for a homogeneous and isotropic distribution of lenses and sources and a distribution of plasma scattering angles that is either a powerlaw (PL) function $P_{\rm scat}=(\theta_{\rm scat}/\theta_{\rm scat,0})^{a}\Theta(\theta_{\rm scat,0}-\theta_{\rm scat})$ or a log-normal distribution $P_{\rm scat}=\mbox{Lognormal}(\log(\theta_{\rm scat,0}),0.5)$.}
		\label{fig:tauvsM}
	\end{figure}

	Combining the conditions given by eqns. \ref{eq:thsmu}, \ref{eq:thscatmu} we can write an expression for the optical depth
	\begin{eqnarray}
		\label{eq:tauscat}
		&   \tau(>\mu)=\frac{32\tau_0}{\theta_{\rm E,0}^2}\int_0^1 dx \frac{dP_{\rm x}}{dx} P_{\rm s}(\theta_{\rm s,\mu}) P_{\rm scat}(\theta_{\rm scat,\mu}) \nonumber \\
		& =\frac{6\tau_0}{\mu^2} \int_0^1 dx\, x(1-x)P_{\rm scat}\left(\frac{\theta_{\rm E,0}}{(1-x)^{1/2}x^{1/2} 4\mu}\right).
	\end{eqnarray}
	In particular we see that in the limit of no scattering, $P_{\rm scat}(\theta\to 0)=1$, $\tau(>\mu)=\tau_0/\mu^2$ as required. The effects of scattering can be understood by examining a couple of concrete examples for $P_{\rm scat}$. First, consider that all plasma screens have the same scattering angle, $\theta_{\rm scat,0}$. In this situation, $P_{\rm scat}=\Theta (\theta_{\rm scat}-\theta_{\rm scat,0})$; where $\Theta$ is the step function.  $P_{\rm scat}$ will equal 1 as long as $x(1-x)<(\theta_{\rm E,0}/4\mu \theta_{\rm scat,0})^2$ and 0 otherwise. Stated differently, there is a critical magnification
	\begin{equation}
		\mu_{\rm scat}=\frac{\theta_{\rm E,0}}{2\theta_{\rm scat,0}}
	\end{equation}
	such that for $\mu<\mu_{\rm scat}$ the optical depth is unaffected by scattering. For $\mu\gg \mu_{\rm scat}$, the contribution to eq. \ref{eq:tauscat} comes from two separate regions of $x$ satisfying $0<x<x_{\rm -}$ and $x_{\rm +}<x<1$, where $x_{-,+}=0.5\pm 0.5\sqrt{1-(\mu_{\rm scat}/\mu)^2}$. The integral in eq. \ref{eq:tauscat} is symmetric in $x$ around $x=0.5$, so it is sufficient to work out the scaling for one of those regions to get $\tau(\mu\gg \mu_{\rm scat})$. For $\mu\gg \mu_{\rm scat}$, $x_{-}\approx 0.5 (\mu/\mu_{\rm scat})^2\ll 1$. Plugging this back to \ref{eq:tauscat}, we see that $\tau(>\mu)\propto \mu^{-2} x_{-}^2\propto \mu^{-6}$. 
	Overall, we have
	\begin{eqnarray}
		\label{eq:taudelta}
		\tau(>\mu)\approx \tau_0  \left\{ \begin{array}{ll}\mu^{-2} & \mu<\mu_{\rm scat} ,\\ \\
			\mu_{\rm scat}^4\mu^{-6} & \mu>\mu_{\rm scat}\ .
		\end{array} \right.
	\end{eqnarray}
	The implication is that even if all plasma screens provide the same scattering angle, the magnification reduces only as a powerlaw function beyond the critical magnification. This is because there is always a small region of space where the lens is sufficiently close to the source or the observer, and the effective scattering angle is small enough so that the magnification is not suppressed by plasma.
	If, instead, the plasma screen scattering angles are distributed as a powerlaw, $P_{\rm scat}=(\theta_{\rm scat}/\theta_{\rm scat,0})^{a}\Theta(\theta_{\rm scat,0}-\theta_{\rm scat})$ with $a>0$, then for $\mu\gg \mu_{\rm scat}$, there is a limiting value of $\tilde{\theta}_{\rm scat}=\theta_{\rm E,0}/2\mu$. For $\theta<\tilde{\theta}_{\rm scat}$, magnification is unsuppressed by plasma for any $x$. As a result, $\tau(>\mu)\propto \mu^{-2} \tilde{\theta}_{\rm scat}^{a}\propto \mu^{-2-a}$. This scaling is relevant as long as contributions from $\tilde{\theta}_{\rm scat}$ are dominant over those from $\theta_{\rm scat,0}$. Overall,
	\begin{eqnarray}
		\tau(>\mu)\approx \tau_0  \left\{ \begin{array}{ll}\mu^{-2} & \mu<\mu_{\rm scat} ,\\ \\
			\max(\mu_{\rm scat}^{-2}\left(\frac{\mu_{\rm scat}}{\mu}\right)^{2+a},\mu_{\rm scat}^4\mu^{-6}) & \mu>\mu_{\rm scat}\ ,
		\end{array} \right.
	\end{eqnarray}
	These results are presented in figure \ref{fig:tauvsM} where we show $\tau(>\mu)$ for different $P_{\rm scat}$distributions.
	
	\subsection{Effect of plasma screen on lens delays and FRB lightcurve}
	\label{time-delay}
	
	For the calculation of arrival time of photons along different trajectories we go back to equation \ref{path-integral}. The exponent in that equation is the phase of the photon at the observer location or the product of its arrival time and frequency. The image location is given by stationary points of the phase, i.e.
	\begin{equation}
		{2\pi (\vec{\theta}_I - \vec{\theta}_s)\over \theta_F^2} - {2\omega R_{\rm s} \vec{\theta}_I\over c\theta_{\rm I}^2 } + \omega {\partial \delta t_p\over \partial \vec{\theta} } = 0,
	\end{equation}
	and the unscattered image location by
	\begin{equation}
		{\vec{\theta}_I - \vec{\theta}_s\over \theta_F^2} = {2\nu R_{\rm s} \vec{\theta}_I\over c\theta_{\rm I}^2 } \;\;{\rm or}\;\; {|\vec{\theta}_I - \vec{\theta}_s|\over \theta_{\rm E}^2} = {1\over \theta_{\rm I}}.
		\label{thetaI1}
	\end{equation}
	Thus, the arrival time of photons associated with the image is
	\begin{equation}
		t = {\pi |\vec\theta_{\rm I} - \vec\theta_{\rm S}|^2\over \omega\theta_F^2} -  \psi(\vec\theta_{\rm I}) + \delta t_p(\vec\theta_{\rm I}) = {2R_{\rm s}\over c} \left[ {\theta_{\rm E}^2 \over 2\theta_{\rm I}^2} - \ln \theta_{\rm I}\right] + \delta t_p(\theta_{\rm I}),
	\end{equation}
	where eqns. \ref{gravity-delay} \&  \ref{thetaI1} were used to obtain the second equality and we have assumed that the plasma causes a small correction to the image location.
	
	Finally, the time difference between the arrival of photons for the two images at angles $\theta_{I1}$ \& $\theta_{I2}$ is 
	\begin{equation}
		\begin{aligned}
			\Delta t & = {2R_{\rm s}\over c} \left[ {\theta_{\rm E}^2 \over 2\theta_{I1}^2} - {\theta_{\rm E}^2 \over 2\theta_{I2}^2} + \ln\left( {\theta_{I2}\over\theta_{I1}}\right)\right] + \delta t_p(\theta_{I1}) - \delta t_p(\theta_{I2}) \\
			& = {2R_{\rm s}\over c} \left[ {\theta_{\rm S} (\theta_{\rm S}^2 + 4\theta_{\rm E}^2)^{1/2}\over 2\theta_{\rm E}^2} + \ln\left( { \sqrt{1 + 4\theta_{\rm E}^2/\theta_{\rm S}^2} + 1 \over \sqrt{1 + 4\theta_{\rm E}^2/\theta_{\rm S}^2} - 1}\right)\right] \\
			& {\hskip 4.5 cm} + \delta t_p(\theta_{I1}) - \delta t_p(\theta_{I2}) 
		\end{aligned}
		\label{Deltfull}
	\end{equation}
	For $\theta_{\rm S} < \theta_{\rm E}$, this reduces to
	\begin{equation}
		\Delta t \approx {4 R_{\rm s} \theta_{\rm S} \over c\theta_{\rm E}} + \delta t_p(\theta_{I1}) - \delta t_p(\theta_{I2})
		\label{delay1}
	\end{equation}
	
	When $\theta_{\rm S} = 2\theta_{\rm E}$, the magnification factor in the absence of plasma for one of the lens images is 1.03 whereas the other image is demagnified by a factor 33; for a source at $3\theta_{\rm E}$, the second image is demagnified by a factor 10$^2$. Thus, identifying superposition of two lens images in the lightcurve of an FRB would be difficult when $\theta_{\rm S} \gta 2\theta_{\rm E}$. We, therefore, provide estimates for the particular case of $\theta_{\rm S}\sim \theta_{\rm E}$, which has the highest probability of occurrence. The geometrical plus the gravitational time difference between the two images in this particular case is $4.16 R_{\rm s}/c$ which is close to the value in eq. (\ref{delay1}). Next we consider the time difference due to wave propagation through turbulent plasma.
	
	The extra time it takes for radio waves of frequency $\omega$ to travel through an eddy of size $\ell$ is $(\ell/c)\omega_p^2/\omega^2$. For uncorrelated eddies, the total extra time to travel through a plasma screen of thickness $L$ is 
	\begin{equation}
		\Delta t_p(\ell) = {\ell\over c} {\delta n_{\rm e}(\ell) \over n_0} \left({L\over\ell}\right)^{1\over 2} {4\pi q^2 n_0 \over m \omega^2} = {(\rm 4.3\, ms) \, DM\over \nu_9^2} \left({\ell\over L}\right)^{1\over 2} \left({\ell\over\ell_{\rm max}}\right)^\alpha
		\label{dtp7}
	\end{equation}
	
	where $\mbox{DM} = L n_0$ is the mean dispersion measure (DM) of the plasma screen. 
	
	The eddy that contributes most to the travel time difference between the two image trajectories has a size of order the distance between the two trajectories in the plasma-plane, i.e. 
	\begin{equation}
		\ell_{12}\sim d_{\rm LO} (\theta^I_+ - \theta^I_-)\sim d_{\rm LO} \left( \theta_{\rm S}^2 + 4 \theta_{\rm E}^2\right)^{1/2} \sim 2 \theta_{\rm E} d_{\rm LO}.
	\end{equation}
	Substituting $\ell = \ell_{12}$ into \ref{dtp7} gives the arrival time difference for the two lensed images due to the presence of turbulent plasma,
	\begin{equation}
		\Delta t_p \sim ({\rm 2\, s)\, DM}\, \nu_9^{-2}\,M_\odot^{5/12}  {[d_{\rm LO} (d_{\rm SO} - d_{\rm LO})]^{5/12}\over d_{\rm SO}^{5/12} L^{1/2}\ell_{\rm max}^{1/3} }
		\label{dtp9}
	\end{equation}
	
	Taking $\ell_{\rm max}\sim 10^{17}$cm, $L=10^2$\,pc, and $d_{\rm LO}= 1$ Mpc, we find from the above equation that $\Delta t_p\sim 0.4$ ms DM$_2 M_\odot^{5/12}$. This should be compared with the geometric + gravitational delay for the two images, $\Delta t_{g}\sim 4 R_{\rm s}/c$ when $\theta_{\rm S}\sim\theta_{\rm E}$ (see eq. \ref{delay1}). For $\Delta t_p\gta \Delta t_g$, the plasma delay dominates the delay between arrivals of radio signals for the two lens images. In this regime, (provided that pulse broadening by scattering is sub-dominant, see \S \ref{sec:magnification} for the appropriate condition) an observer would see two pulses with similar spectro-temporal evolution (after de-dispersing the signal) and with slightly different DM values (note that $\Delta \mbox{DM}(\ell)/\mbox{DM}\sim \Delta t_p(\ell)/t_p<1$). This  will allow to establish that the lightcurve is a superposition of two lens images. However, the time delay in this regime is no longer a direct proxy for the lens mass. In particular, the connection between magnification and time-delay that one calculates for gravitational lensing from a point source \footnote{This connection is found by relating the gravitational + geometric components of $\Delta t$ in eq. \ref{Deltfull} to the relative magnification of the two images (eq. \ref{Mgl}). The result (when plasma scattering can be ignored) is $\Delta t=2GM_{\rm l}\left(\frac{\gamma-1}{\gamma}+\log \gamma \right) / c^3$ where $\gamma$ is the ratio of magnifications of the two images (see e.g. \citealt{Yang2021}).} will be modified. This consideration places a lower limit on lens masses above which plasma-delay is insignificant,
	\begin{equation}
		M_{\rm min,t} \sim 0.05\, M_\odot\, {\mbox{DM}_2^{12\over 7} {\rm min}\big\{d_{\rm LO,22}, d_{\rm SL,22}\big\}^{5\over7}\over \nu_9^{24\over 7} L_{22}^{6 \over7} \ell_{\rm max,17}^{4\over7} }.
		\label{M-lens-min}
	\end{equation} 
	Surveys at high frequencies can explore smaller lens masses as $M_{\rm min,t} \propto \nu^{-24/7}$.

	The scenario we have described thus far considers the lens to be located inside a plasma screen. However, the calculation can be easily extended to a wider set of possibilities. For instance, when the plasma density in the lens-plane is small but the photon trajectories corresponding to the two images pass through a turbulent plasma screen somewhere between the source and the observer, equation (\ref{dtp7}) can be used for calculating the travel time difference along different photon trajectories by substituting $\ell\sim 2\theta_{\rm E} d_{\rm LO} \times f_d$. The factor $f_d= d_{\rm PO}/d_{\rm LO}$ when the plasma screen is between the lens and the observer, and $f_d = d_{\rm SP}/d_{\rm SL}$ when the plasma lies between the source and the lens; where $d_{\rm PO}$ ($d_{\rm SP})$ is the distance between observer (source) and the plasma screen. The factor $f_d$ accounts for the fact that the distance between the photon trajectories for the two images in the plasma plane is smaller than   $\sim 2R_{\rm E}= 2\theta_{\rm E} d_{\rm LO}$ by the factor $f_d^{-1}$. 
	
	The effect of plasma on micro-lensing in the FRB host galaxy is similar to the effect we have 
	discussed above.
	
	\subsubsection{IGM turbulence}
	The size of the largest eddy in the IGM, $\ell_{\rm max}$, is highly uncertain by several orders of magnitude. It could be as large as 10$^{24}$cm --  the scale for energy deposition into the IGM by AGN jets and outflows from galaxy clusters -- or as small as a 10$^{20}$cm. If the Mach number of the IGM turbulence were to be $\xi_{m}$ then $\ell_{\rm max} \gta 10^{24} \xi_{m}^3$, otherwise the heating of the IGM by dissipation of turbulent energy will raise its temperature on a time scale smaller than the Hubble time, which is contradicted by observations; the IGM data show that the mean temperature is $\sim 10^4$K and is not increasing with decreasing redshift.
	
	Therefore, taking $\ell_{\rm max}\sim 10^{22}$cm, $L\sim d_{\rm LO}$, $d_{\rm SL}\sim 10^{28}$ cm and ${\rm DM}\sim 10^3$ pc cm$^{-3}$ for the IGM turbulence, we find from eq. (\ref{M-lens-min}) that the time delay between two images of an object due to IGM turbulence is larger than the gravity+geometry effects for lens mass $\lta 10^{-3} M_\odot$. We note that according to our estimates, scattering of radio waves by IGM turbulence is much weaker than expected for FRB host galaxy and the Milky Way ISM, which is consistent with \cite{Cordes2022} \& \cite{Ocker2022}.
	
	\section{Synergistic effects of plasma scintillation and gravitational lensing} 
	So far we have focused on ways in which plasma scattering suppresses the magnification, smears out lightcurves of the two images, and changes time delays of gravitational lensing events. These ultimately limit the range of parameter space in which FRB lensing can be observed and used for cosmology. In this section, we show that plasma scattering combined with gravitational lensing can also, under certain circumstances, lead to unique constraints on cosmology or on the nature of the environment of the FRB source.
	
	\subsection{Using FRB lensing delays to constrain cosmology}
	\label{sec:cosmologyapp}
	
	Consider a situation in which $\theta_{\rm scat}'\ll \theta_{\rm E}$ and $t_{\rm sc}\sim(2R_{\rm s}/c)(\theta_{\rm scat}'/\theta_{\rm E})^2< t_{\rm FRB}<\Delta t_{g}+\Delta t_p$. As described in \S \ref{sec:magnification}, \S \ref{time-delay}, in this situation the FRB lightcurve will appear as consisting of two distinct and similarly shaped components due to the lensing as the time delay between the images is long compared to the intrinsic FRB duration which in turn is longer than the scintillation broadening timescale (we have assumed $\theta_{\rm s}\approx \theta_{\rm E}$ in the expression above for clarity and since this is the most relevant case of interest). The plasma broadening for each component of the lightcurve corresponding to an image may still be detectable. Furthermore, the temporal broadening of each component will be slightly different due to the different columns of turbulent eddies along the photon trajectories for the two images and the slightly different DMs along the two paths. 
	Indeed, estimating the DM separately for the two components of the FRB lightcurve will provide us with an estimate of $\Delta \mbox{DM}(\ell)$ along the two photon trajectories that are separated by a distance $\ell$. The physical separation between the two paths, in the lens plane, is typically of order $\sim R_{\rm E}$. As an example, for a $10^{11}M_{\odot}$ lens at a cosmological distance, we can measure $\Delta \mbox{DM}$ on a scale of $5\mbox{ kpc}$. 
	This corresponds to a tiny angular separation, $\Delta \theta \sim 10^{-6}$ rad. So, lensing could be used to probe density fluctuations on this small length scale, which cannot be done by two unrelated FRBs (because of their large typical angular separation) or any other means currently available to us. Lensing, therefore, provides us with a unique way to constrain the fluctuations of $\mbox{DM}_{\rm IGM}$, and explore the patchy reionization at $z>6$. Moreover, lensing of a repeating FRB might be useful for probing the time dependence of density fluctuation. It should be noted that the size of ionizing bubbles at the end of the reionization epoch is much larger, $\sim 10$\,Mpc \citep{2004Natur.432..194W}, so only minor fluctuations in $\mbox{DM}_{\rm IGM}(z)$ are expected on the angular scales that are probed by lensing. Still, given that at large redshift, $\mbox{DM}(z\gtrsim 6)\approx6000\mbox{pc cm}^{-3}$ \cite{Beniamini+21}, even very small fractional changes in $\mbox{DM}_{\rm IGM}$ may still be detectable.

	\begin{figure}
		\centering
		\includegraphics[width = 0.25\textwidth]{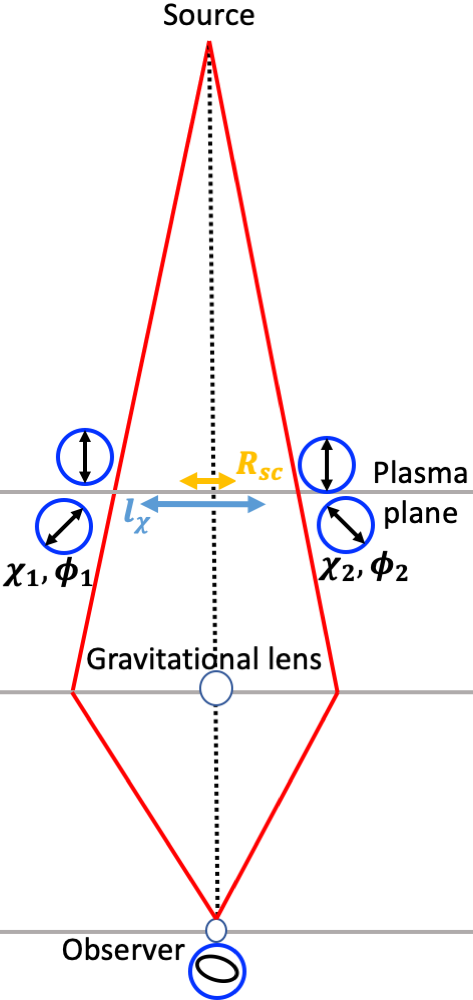}
		\caption{ Schematic figure demonstrating how multi-path propagation by gravitational micro-lensing leads to different parts of the FRB wave accumulating different phases ($\phi_i$) and rotation ($\chi_i$) of their electric wave vector. In this setup, the resulting signal, which is a superposition of the images, will generally be elliptically polarized, even if the magnetic field changes on a spatial scale ($\ell_{\chi}$) that is large compared to the visible size of the plasma screen ($R_{\rm sc}$) in the absence of the gravitational lens.}
		\label{fig:schematic2}
	\end{figure}
	
	\begin{figure*}
		\centering
		\includegraphics[width = 0.42\textwidth]{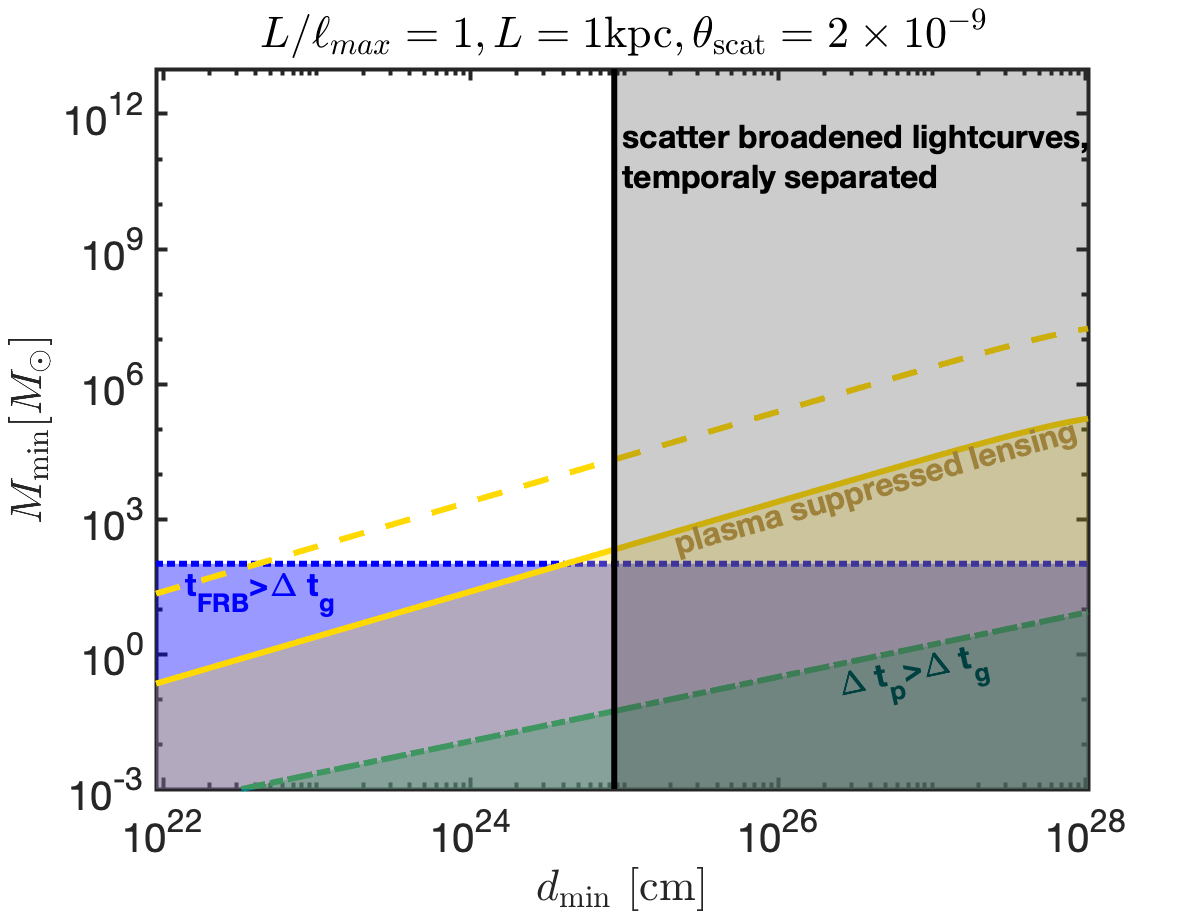}
		\includegraphics[width = 0.42\textwidth]{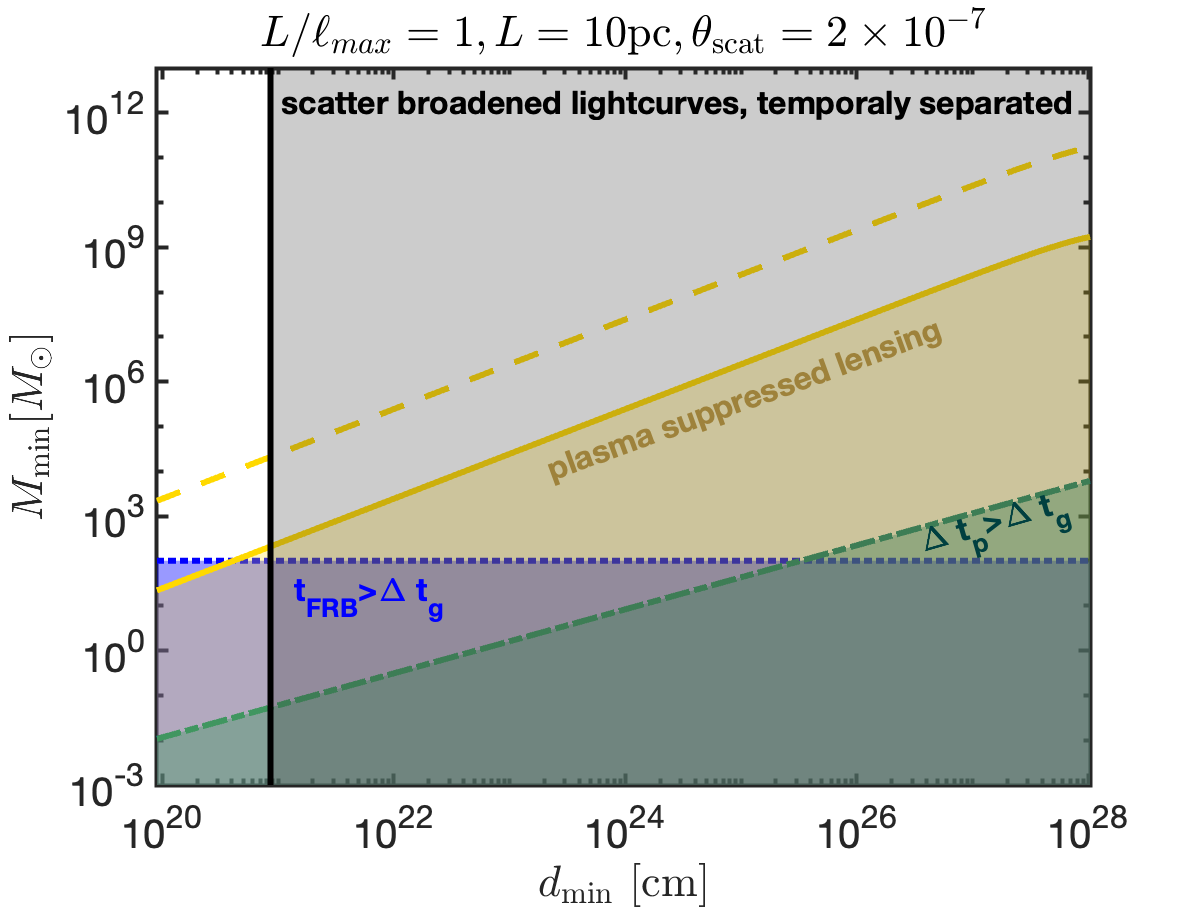}\\
		\includegraphics[width = 0.42\textwidth]{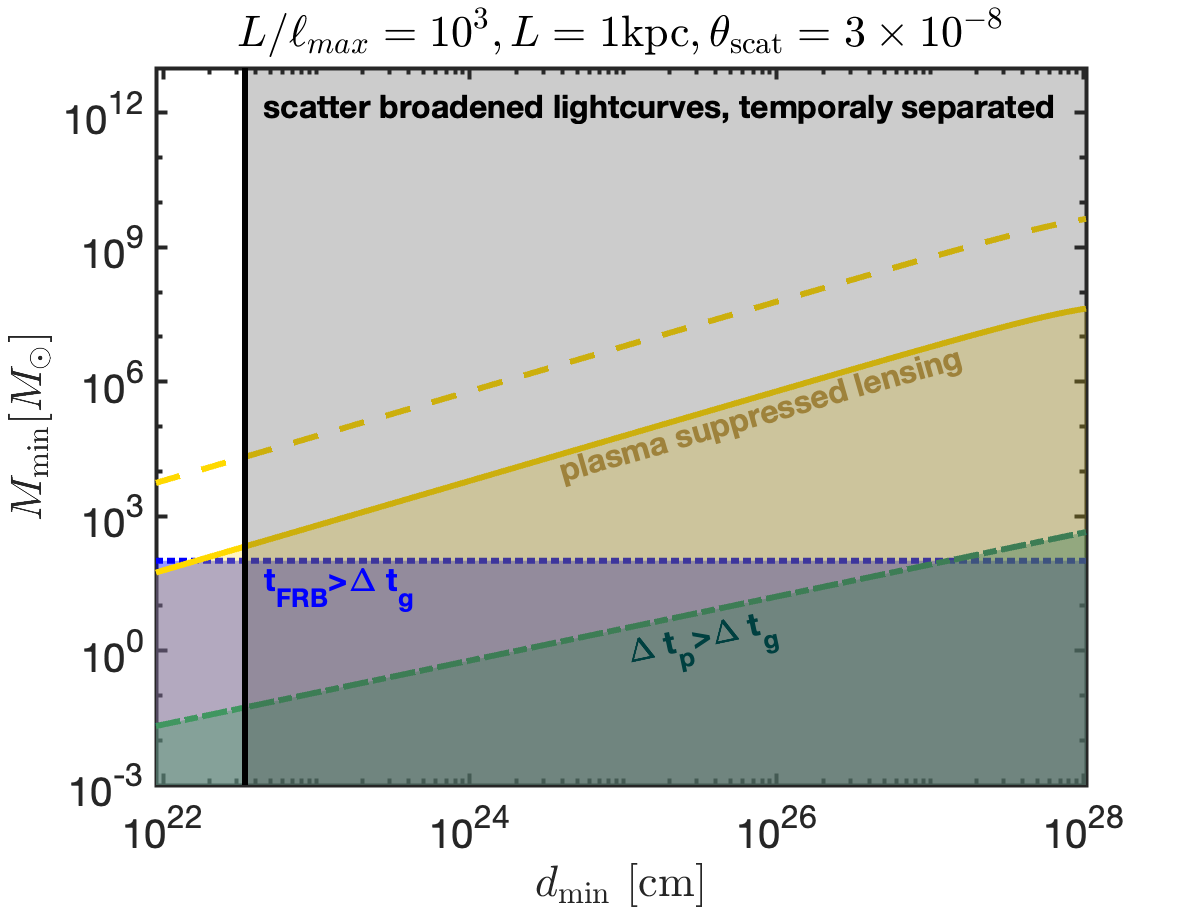}
		\includegraphics[width = 0.42\textwidth]{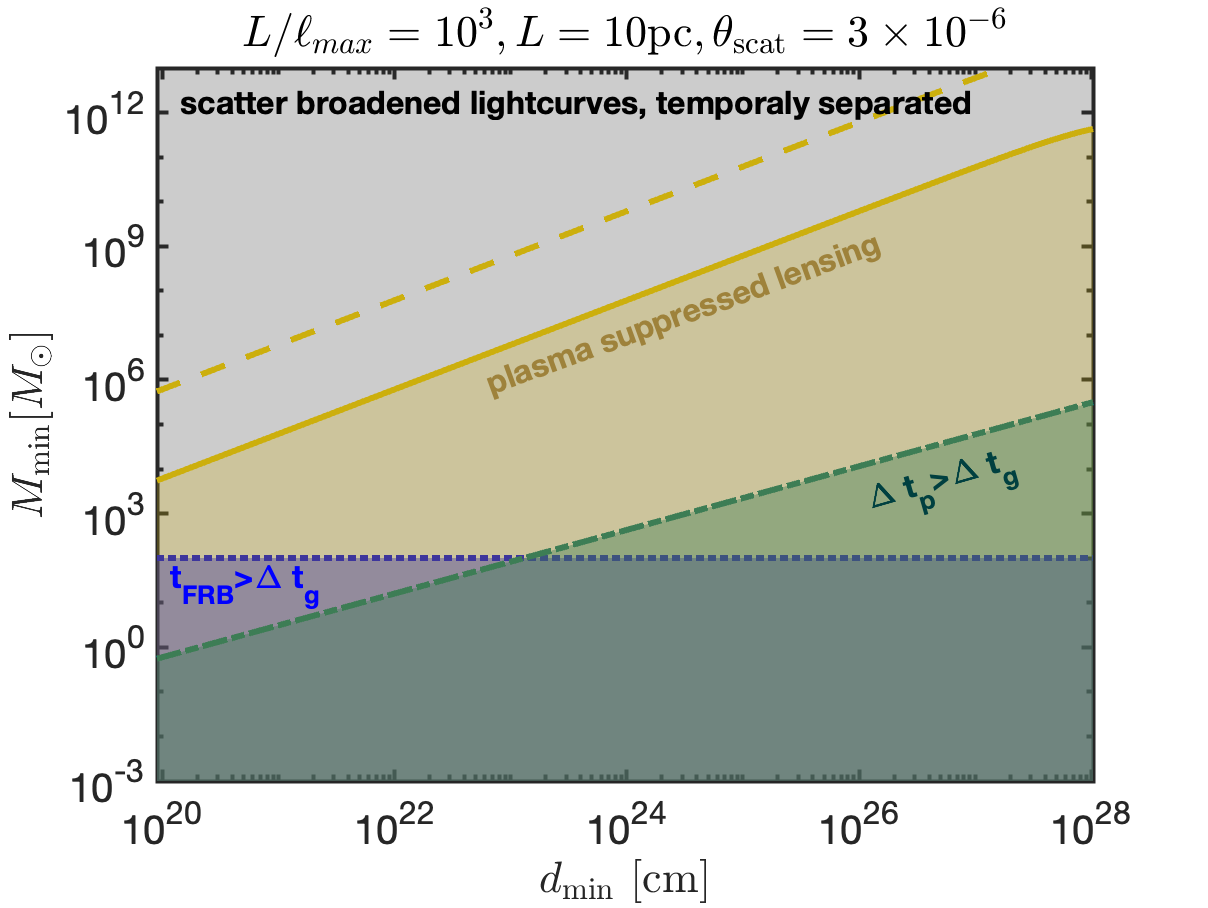}
		\caption{Minimum lens masses that can be probed with FRB lensing. A detectable time delay between the separate images of a lensed FRB and a magnification above unity, can only be seen above the diagonal solid line. Larger magnifications require greater lens mass to overcome the blurring effect of turbulence ($M_{\rm min,\mu}\propto \mu_{\rm max}^2$, see eq. \ref{eq:Mlimmag2}). $M_{\rm min,\mu}(\mu_{\rm max}=10)$ is shown by a dashed line. A vertical line denotes the value for $\min(d_{\rm SL},d_{\rm LO})$ above which the lensed images become scatter broadened. The temporal delay between images of a lensed FRB is dominated by the geometrical + gravitational delay ($\Delta t_g$), rather than the plasma delay, above the dot-dashed line. $\Delta t_g$ is larger than the intrinsic FRB duration ($t_{\rm FRB}$) above the dotted line. These plots demonstrate that the limiting criteria for detection of FRB lensing are typically given by $M_{\rm min,\mu}$. Results are plotted as a function of $d_{\rm min}=\min(d_{\rm LO},d_{\rm SL})$ and assuming that the plasma screen is in the vicinity of the gravitational lens. Other parameters assumed for plotting these figures are: $d_{\rm SO}=2\cdot 10^{28}\mbox{cm},\mbox{DM}=10^2\mbox{pc cm}^{-3}, \mu_{\rm max}=1, \nu=1\mbox{GHz}, t_{\rm FRB}=1\mbox{ ms}$. Different panels represent different values of $L, L/\ell_{\rm max}$ (corresponding to different values of $\theta_{\rm scat}$ listed in each panel title); where $L$ is the width of the plasma screen \& $\ell_{\rm max}$ is the size of largest eddies in the turbulent medium. }
		\label{fig:Mlim}
	\end{figure*}
	
	\subsection{Induced circular polarization in the presence of gravitational lensing}
	\label{sec:circular}
	A radio burst passing through a plasma with inhomogeneous density and magnetic field strength, can become partially circularly polarized and possibly also depolarized when it reaches the observer \citep{BKN2022}. The circular polarization stems from the fact that rays reaching the observer at a given time, have propagated through different segments of the plasma and by doing so have accumulated different phases and different degrees of Faraday rotation.
	For a plasma screen, this effect is important when $\ell_{\chi}<R_{\rm sc}$, where $\ell_{\chi}$ is the separation (along the plane of the plasma screen) over which the difference in rotation between two waves going through the screen is of order unity and $R_{\rm sc}$ is the scattering radius, which is the visible size of the screen.
	
	Gravitational micro-lensing can enhance the effect of an inhomogeneous plasma screen if the projection of the Einstein radius on the plasma screen is, $R_{\rm E} f_d>R_{\rm sc}$ (where we have assumed here that the distance between the images in the lens plane is $\sim R_{\rm E}$ for the case in which lensing causes significant magnification). In this case, assuming Kolmogoroff turbulence, the condition for induced circular polarization by multi-path propagation becomes
	\begin{eqnarray}
		\label{eq:cirpol}
		&	f_d \frac{R_{\rm E}}{\ell_{\chi}}>1\rightarrow \nonumber \\
		&	110\frac{f_d\mbox{RM}_{4}^{6/5}(M/100M_{\odot})^{1/2}\min\{d_{\rm LO,28},d_{\rm SL,28}\}^{1/2}}{\nu_9^{12/5}\ell_{\rm max,18}^{2/5}L_{20}^{3/5}}>1
	\end{eqnarray}
	where RM is measured in units of $\mbox{rad m}^{-2}$.
	Eq. \ref{eq:cirpol} shows that if (for example) the plasma screen is in the vicinity of the gravitational lens, then gravitational micro-lensing can enhance the induced circular polarization. An additional requirement for this to occur is that the lens should cause significant magnification of the source, as described in \S \ref{sec:magnification} and figure \ref{fig:Mscatter} (such that there are multiple images that contribute significantly to the measured flux). The degree of induced circular polarization is of order unity in case this inequality is satisfied, and of order the L.H.S. otherwise.
	
	Having multiple paths due to gravitational (rather than plasma) lensing means that it is not essential for the magnetic field in the different paths to be originating from a single, strongly turbulent plasma screen. Instead, different rays may be intersecting plasma with different properties. In order for the degree of rotation to significantly change between two such paths, we require that
	\begin{equation}
		\label{eq:minB}
		\Delta \chi=0.07\nu_{9}^{-2}\left(B_{\mu G}\Delta \mbox{DM}+\Delta B_{\mu G}\mbox{DM}\right)>1
	\end{equation}
	At the same time, light from the two paths must interfere, meaning that
	\begin{equation}
		\label{eq:maxDM}
		\Delta t_p+ \Delta t_g<t_{\rm FRB}\sim 1\mbox{ ms}\rightarrow \Delta \mbox{DM}<0.23\nu_9^{2}\left(1-\Delta t_g/\mbox{ ms}\right)
	\end{equation}
	It is possible to simultaneously satisfy both eqs. \ref{eq:minB} and \ref{eq:maxDM} with astrophysically plausible parameters. For example, if the lens is a stellar mass black hole binary losing mass through wind at the Eddington rate, its wind may extend up to $\sim 0.1$\, pc. A ray intersecting this wind at $\sim 10^{16}$\,cm (slightly less than the typical Einstein radius for such a black hole), would acquire an excess DM of order $\Delta \mbox{DM}\sim 10^{-3}\mbox{pc cm}^{-3}$ (for a $10M_{\odot}$ black hole). At the same distance, the wind has a magnetic field strength of order $B\sim 10\mbox{ mG}\sigma_{-2}^{1/2}$ where $\sigma$ is the magnetization parameter. For these parameters we have $\Delta \chi \sim 1 \nu_9^{-2}$, demonstrating that the conditions \ref{eq:minB} and \ref{eq:maxDM} are satisfied at  $\nu\lesssim 10^9$\,Hz and that this setup would result in strong circular polarization.

	\section{Conclusions and discussion}
	\label{sec:conc}
	We have investigated in this work how gravitational lensing of point radio sources, such as fast radio bursts (FRBs), by a point mass is affected by a plasma screen between the source and the observer. The main results we found are summarized and discussed below. 
	
	Much of the analysis in this work was presented for the case in which the plasma screen is co-located with the gravitational lens. However, gravitational lensing by stars is likely to preferentially occur in old and massive elliptical galaxies. These galaxies have a more tenuous ISM, which reduces the chance for the radio waves encountering a plasma scattering screen within the same galaxies. This motivates us to consider the extent to which plasma scattering in the FRB host galaxy or the Milky Way can affect gravitational lensing.
	When the plasma screen is at a general location relative to the gravitational lens, and the scattering angle for photons through the plasma is $\theta_{\rm scat}$, the image magnification is capped at $\theta_{\rm E}/2\theta_{\rm scat}'$; where $\theta_{\rm scat}'=\theta_{\rm scat} f_d d_{\rm SL}/d_{\rm SO}$, $\mbox{S,L,O,P}$ stand for the source, lens, observer, and plasma screen, $d_{\rm XY}$ is the distance between $X$ and $Y$ and $f_d=d_{\rm PO}/d_{\rm LO}$ ($f_d=d_{\rm SP}/d_{\rm SL}$) when the plasma is between the observer and the lens (source and lens). This effectively translates to a lower limit on the lens mass, $M_{\rm min,\mu}$ (eq. \ref{eq:Mlimmag2}) that can be probed with FRB lensing. For masses below $M_{\rm min,\mu}(\mu_{\rm max}=1)$, the presence of gravitational lensing cannot be inferred from either the magnification or by spotting a  duplicate copy of the signal in the lightcurve that is delayed w.r.t. the first component. In particular, we note that $M_{\rm min,\mu}\propto f_d^{-2}$, and that means that plasma effects on lensing magnification and pulse broadening are strongly suppressed for gravitational lenses at cosmological distances when plasma scattering takes place far away from the lens either in the FRB host galaxy or in the Milky Way.
	
	As a specific example for the role of plasma suppressed gravitational lensing, consider the case of FRB 20191221A. This bright FRB is particularly remarkable due to its long duration ($\sim 3$\,s), and 217 ms periodicity \citep{subsecondP2022}. One possibility explored by the authors to explain the uniqueness of this burst is that it could be an ordinary extra-galactic pulsar that has been micro-lensed by its binary companion and the flux amplified by a factor of $\sim 10^{11}$. \cite{subsecondP2022} pointed out the highly unlikely geometry required for this tremendous magnification. The problem is made much worse by non-zero scintillation in the host galaxy, which limits the lens magnification ($\mu$) to $\sim 10^5$. The lensing model requires $d_{\rm SL}\sim 10\mbox{pc}, d_{\rm LO}\sim 1\,\mbox{Gpc}, {\rm and}\, M_{\rm l}>10^6M_{\odot}$. The Einstein angle for these parameters is $\theta_E\sim 10^{-12}$rad. If the scattering angle for the plasma in the host galaxy were to be similar to the Milky Way galaxy, i.e. $\theta_{\rm scat} \sim 10^{-9}$rad or $\theta'_{\rm scat}\sim 10^{-17}$rad, then the maximum magnification is limited to $\theta_E/2\theta'_{\rm scat}\sim 10^5$ due to broadening of the angular size of the source. This is much smaller than the required $\mu \sim 10^{11}$. The required magnification can be achieved provided that $M_{\rm l}\gta 7\cdot 10^{17}\theta_{\rm scat,-9}^2$ (see eq. \ref{eq:Mlimmag}), which is unphysical even when we take into account the uncertainty in the value of $\theta_{\rm scat}$.
	
	The lensing probability, $\tau(>\mu)$, is modified by scintillating plasma as discussed in \S\ref{sec:tau}. The probability is suppressed for high magnification events as waves scattered by the turbulent medium between the source and the observer increase the angular size of the source to $\sim \theta'_{\rm scat}$, and this reduces the magnification. A corollary of this is that $\tau(>\mu)$ is no longer proportional to $\mu^{-2}$ for $\mu\gg 1$, but falls off more steeply. 
	
	Different dispersion measure (DM) along photon trajectories for the two different images introduces an extra time delay between the two images of a transient source. The distance between the photon trajectories in the lens plane is of order the Einstein radius ($R_E$), and the extra time delay ($\Delta t_p$) is proportional to $R_E^{5/6}$ due to waves traveling through a turbulent medium with Kolmogoroff spectrum for density fluctuation.
	The effect is most severe for stellar and sub-stellar mass lens when $\Delta t_p$ is comparable to the gravitational time delay or the duration of FRBs. The implication of this plasma introduced extra time delay between the two copies of an FRB is that the observed delay is no longer a direct proxy for the lens mass, and that the plasma delay must be corrected for determining the cosmic abundance of stellar mass dark matter. At the same time, the fact that $|\Delta t_p|>0$, can under certain circumstances, also be an advantage. When $\theta_{\rm scat}'\ll\theta_{\rm E} \mbox{ \&  }\Delta t_g + \Delta t_p> t_{\rm FRB}$, the existence of lensing can be inferred from the lightcurve, while the different DM along the different image trajectories allow us to measure $\Delta \mbox{DM}_{\rm IGM}$ on tiny angular scales, $\lesssim 10^{-6}$\,rad, which cannot be explored by any other observational means. This technique might also turn out to be useful in the study of patchy reionization history of the universe.
	
	Gravitational lensing can convert a linearly polarized source to partial circular polarization when the time delay between the images is less than the coherence time for the source. The reason for this is that photons traveling through a magnetized plasma suffer different amounts of rotation of the electric vector (different rotation measures, RM), and different phase shifts, along the two image paths, thereby resulting in some degree of circular polarization.
	
	There is a lens mass above which image magnification and time delays are not affected much by the turbulent plasma between the source and observer. This minimum lens mass is shown in figure \ref{fig:Mlim} for a few different combinations of length scales associated with the plasma screen and the distance of the plasma screen and the lens from the source/observer. Lensing by $\lesssim 10^2M_{\odot}$ objects may be strongly suppressed by plasma scattering and the pulses of the lensed images become temporally overlapping. That makes it hard to spot gravitational lensing of FRBs without resorting to specialized analysis that can accurately remove different scatter-broadening of the two images in the lightcurve. An additional consideration that the relative plasma delay for the two images should be short compared with the geometrical + gravitational delay turns out to be easier to satisfy, as it typically kicks in at lower lens mass than quoted above.
	The third constraint is that the scatter broadening timescale, $t_{\rm sc}$, should be short compared to $t_{\rm FRB}$ in order that the two lensed images have a similar temporal profile so that observers could identify the lensing event\footnote{It might be possible to identify an FRB lensing event even when the temporal profiles of the images look very different due to scattering by turbulent plasma along the different photon trajectories of the images by e.g. a cross-correlation analysis that deconvolves the different scatter broadening for different images.}. This is satisfied when $\min(d_{\rm SL},d_{\rm LO})$ is smaller than a critical distance, $d_{\rm sc}$ (when the plasma plane is separated from the lens plane, the condition becomes $\min(d_{\rm SP},d_{\rm PO})<d_{\rm sc}$); where $d_{\rm sc}$ is given by eqns. \ref{eq:dsc} \& \ref{eq:dsc2}.
	The plasma scattering reduces rapidly with frequency (see eqns. \ref{eq:Mlimmag2} and \ref{M-lens-min}). As a result, lensing of FRBs is easier to observe at higher frequencies. 
	
	The effect of scintillation on gravitational lensing was considered also by \cite{KKSX2020}. They focused on a particular lensing scenario where the lightcurves of the images overlap and produce interference fringes. Much of their work was devoted to the analysis of the effect of this interference on the observed spectrum. By contrast, the present work describes how scintillation affects the magnification, lensing probability, image lightcurves and their time delays. \cite{KKSX2020} analysis is for the regime where $\Delta t_g^{-1}$ is larger than the spectral resolution of the detector, but smaller than the coherence bandwidth of the source, and that translates to lens masses in the range $10^{-4}M_{\odot}<M_{\rm l}<10^{-1}M_{\odot}$ as per these authors.  The ``Time domain" lensing effects discussed in this paper should be applicable to a larger range of lens mass and wider parameter space of scintillating plasma screen.
	Furthermore, the condition $\Delta t_p>\Delta t_g$ is expected to apply for $10^{-4}M_{\odot}<M_{\rm l}<10^{-1}M_{\odot}$ (see figure \ref{fig:Mlim}), and that means that even for the parameter space explored by \cite{KKSX2020}, the determination of lens mass from the lensing time delay signal is affected by the physical effects described in this work.

	\section*{Acknowledgments}
	
	This work has been funded in part by an NSF grant AST-2009619. PB's research was supported by a grant (no. 2020747) from the United States-Israel Binational Science Foundation (BSF), Jerusalem, Israel. Some of the work presented here was carried out while PK was visiting the Yukawa Institute, Kyoto. He is grateful to Kunihito Ioka for his hospitality, for organizing an FRB workshop during that visit, and for many stimulating science discussions. He would like to acknowledge many excellent discussions with Kohta Murase who was also a visitor at the Yukawa Institute. He would like to thank YITP for the  financial support provided under the Visitors' Program of FY2022, and YITP-W-22-18 for funds for the worshop. We are indebted to James Cordes for numerous comments on the draft and for pointing out several previous work on this topic that we were not familiar with. We would also like to thank Bing Zhang, Ue-Li Pen, Liam Connor, Casey Law, Julian Mu{\~n}oz and Marc Kamionkowski for helpful comments on the manuscript.

\end{document}